\RequirePackage{rotating}
\documentclass[acmsmall, screen,nonacm]{acmart}
\usepackage{supertabular}
\usepackage{longtable}
\usepackage{lscape} 
\usepackage{comment}
\usepackage{url}
\usepackage{graphicx}
\usepackage{amsfonts}
\usepackage{color,soul}
\usepackage{multicol}
\usepackage{multirow}
\usepackage{color, colortbl}
\usepackage{blindtext}
\usepackage{array}
\usepackage{tikz}
\usepackage{longtable}
\usepackage{multicol}
\usepackage{gensymb}
\usepackage{caption}
\usepackage{subcaption}

\AtBeginDocument{%
  \providecommand\BibTeX{{%
    \normalfont B\kern-0.5em{\scshape i\kern-0.25em b}\kern-0.8em\TeX}}}

\setcopyright{acmcopyright}
\copyrightyear{2022}
\acmYear{2022}
\acmDOI{XXXXXXX.XXXXXXX}

\acmJournal{JACM}
\acmVolume{00}
\acmNumber{0}
\acmArticle{0}
\acmMonth{0}

\begin{document}

\title{A Survey on Privacy of Personal and Non-Personal Data in B5G/6G Networks}

\author{Chamara Sandeepa}
\authornotemark[1]
\authornote{Manuscript submitted to ACM.}
\email{abeysinghe.sandeepa@ucdconnect.ie}
\author{Bartlomiej Siniarski}
\authornotemark[2]
\email{bartlomiej.siniarski@ucd.ie}

\author{Nicolas Kourtellis}
\authornotemark[3]
\email{nicolas.kourtellis@telefonica.com}

\author{Shen Wang}
\authornotemark[4]
\email{shen.wang@ucd.ie}

\author{Madhusanka Liyanage}
\authornotemark[5]
\email{madhusanka@ucd.ie}
\email{madhusanka.liyanage@oulu.fi}

\renewcommand{\shortauthors}{C. Sandeepa, et al.}

\begin{abstract}
The upcoming Beyond 5G (B5G) and 6G networks are expected to provide enhanced capabilities such as ultra-high data rates, dense connectivity, and high scalability. It opens many possibilities for a new generation of services driven by Artificial Intelligence (AI) and billions of interconnected smart devices. However, with this expected massive upgrade, the privacy of people, organizations, and states is becoming a rising concern. The recent introduction of privacy laws and regulations for personal and non-personal data signals that global awareness is emerging in the current privacy landscape. Yet, many gaps need to be identified in the case of two data types. If not detected, they can lead to significant privacy leakages and attacks that will affect billions of people and organizations who utilize B5G/6G. This survey is a comprehensive study of personal and non-personal data privacy in B5G/6G to identify the current progress and future directions to ensure data privacy. We provide a detailed comparison of the two data types and a set of related privacy goals for B5G/6G. Next, we bring data privacy issues with possible solutions. This paper also provides future directions to preserve personal and non-personal data privacy in future networks. 
\end{abstract}

\ccsdesc[500]{Privacy~Data Privacy}
\ccsdesc[400]{Privacy~Personal and non-personal data}
\ccsdesc[500]{Wireless Networks~Beyond 5G/6G}
\ccsdesc[500]{General and reference~Surveys and overviews}

\keywords{non-personal data, personal data, privacy, beyond 5G/6G}

\maketitle

\section{Introduction}

The fifth generation (5G) of mobile networks is in the phase of deployment and revision, and even though 5G is not implemented in many countries, mobile operators and researchers are already defining the sixth generation (6G) of mobile networks. The novelty and complexity of 6G networks raise new concerns about security and privacy. Adversaries have become more powerful, intelligent, and capable of creating new forms of security threats, especially if considering the pervasive use of AI. Given the importance of data collection and sharing in today's digital and network economies, any misuse of collected data can pose serious risks to users, i.e., the adversary may use information obtained to intimidate or blackmail subscribers. As data privacy compliance requirements for 6G networks become more complicated to manage, these risks will increase. In this section we will discuss the importance of data privacy and present our motivation.
\vspace{-3mm}

\subsection{Importance of Data Privacy In 6G}
For years, data privacy has been a problem partially mitigated by various data protection initiatives, such as the EU Data Protection Directive, Gramn Leach Bliley Act, E-Government Act, EU Right to be Forgotten, and GDPR. The new regulations have driven the advent of privacy preservation techniques such as radio fingerprinting in the physical layer, data and communication anonymization in the connection layer and differential privacy, Homomorphic encryption, data masking, or secure multi-party computing in the service layer. However, there are a number of reasons why privacy protection in 6G is more critical. 

Firstly, in 6G, it is anticipated that key applications such as wearable devices will handle users' more sensitive data. Despite the fact these applications can help to improve human lives by reducing the risk of fatal accidents, promoting restful sleep, and facilitating the rehabilitation of people with disabilities, the physical and medical data for control systems that coordinate these interconnected applications may be illegally gathered and abused. These threats might not be unique, but 6G will amplify them. Secondly, in dense networks, more precise localization through telecommunications is a major concern. The concept of highly accurate access points that track a user's motion with centimeter-level precision to improve link connectivity raises huge concerns as user data can be used for surveillance. Thirdly, the cloudification of many core components and applications in 6G is a flawed strategy. By migrating workloads to the cloud, the risk of unauthorized access and exposure to customer data, including illegal disclosure by unauthorized employees, increases. Finally, securing personal data in the age of supercomputing and intelligent agents is challenging. The need for AI-enabled smart applications is anticipated to increase tremendously with the advent of a 6G network that connects humans and objects. These AI-powered applications can unearth additional context-related information about a particular individual and his or her environment. By utilizing personal or other confidential data, AI can give consumers more precise and intelligent personalized services, such as recommendations of tourist attractions, films, and routes. However, this type of experience also impacts the privacy of consumers. Users may be unaware that they are the targets of large data gathering for unsolicited advertising; far worse would be AI-powered stalking and extortion. 

The privacy-preserving technologies some of which are presented in this survey, by definition allow users to protect the privacy of their personally identifiable information provided to and handled by service providers or applications, all while allowing marketers to maintain the functionality of data-driven systems. The biggest challenge presented by the literature is a trade-off between assuring high privacy protection for individuals with their right to be forgotten and data mining to maximize the accuracy of recommendations/guidelines for users. Particularly in a data-driven industry that is data-hungry, the line between supplying useful information and being exploited for monetary gain is thin.
\vspace{-3mm}

\subsection{Motivation for Our Paper}
When considering the progress of the B5G/6G concept over recent years, the initial phase of identifying core requirements and ideation of B5G/6G has already started. Several surveys aggregate the related works that are on the 6G core network, requirements, architecture, applications, enabling technologies, and the future roadmap~\cite{porambage2021roadmap,zhao2020comprehensive,shahraki2021comprehensive,de2021survey,dogra2020survey,huang2019survey,lu20206g}. These papers do not consider privacy as the main topic but analyse the initial foundation of B5G/6G related works and vision. Some recently published surveys discuss the security and privacy aspects in B5G/6G networks~\cite{sandeepa2022survey,nguyen2021security,wang2020security,sun2020machine}. However, none of them consider aspects of personal and non-personal nature in data. Especially, the non-personal data is not observed for any consideration related to 6G yet, though it possesses a significant impact on privacy that we identified.
Further discussing the related works on privacy and 6G, the survey in~\cite{sandeepa2022survey} provides privacy issues and potential solutions based on a layered architecture for B5G/6G. However, it lacks a discussion relative to personal and non-personal data. The authors in~\cite{nguyen2021security} discuss security aspects and some ideas on privacy in B5G/6G. They discuss privacy matters and challenges, trusted, untrusted, and semi-trusted privacy preservation for 6G. However, the discussion of privacy seems lesser relative to security. Another discussion is in~\cite{wang2020security}, where the authors provide issues related to both security and privacy. Here, they provide issues without making much difference between security and privacy, where specific privacy issues may be difficult to distinguish from security issues. The survey in~\cite{sun2020machine} provides a discussion on Machine Learning (ML) which can be used as a key component in 6G automation. However, ML-related issues are only a part of privacy issues. Also, no significant discussion is found for non-personal data issues or solutions with ML.

As observed, the key motivation for this survey is the lack of consideration of the factors of personal and non-personal data in the related works in B5G/6G. It is a critical requirement to distinguish the nature of data shared within 6G and be regarded as the first step of privacy for any operation or service. Raw data should be able to correctly classify whether they belong to a natural person or not. Then, it will be possible to identify potential loopholes in privacy, such as unmet privacy requirements and gaps in privacy protection. Next, it will be possible to apply possible solutions suited to each problem. Therefore, we present this survey as the starting point for this discussion on the privacy of personal and non-personal data with importance, concerns, and solutions for B5G/6G.

\vspace{-3mm}
\subsection{Our Contribution}
Our survey is a comprehensive study of related literature on personal and non-personal data privacy within the context of B5G/6G. As indicated by Table~\ref{tab:surveys}, we identified the existing relevant surveys do not initiate a sufficient or any discussion towards our topic. They only provide generic privacy ideas, mixed concepts related to security with privacy, or privacy of only specific areas in 6G, which do not essentially have a connection with personal and non-personal data. To the best of our knowledge, this is the first attempt at comprehensively investigating the impact of privacy in non-personal data versus personal data privacy-related aspects of B5G/6G networks. We summarise our contributions in these key points as follows:
\vspace{-1mm}
\begin{itemize}
    \item \textbf{Determine how data privacy has evolved over mobile network generations:} Investigate the evolution of the concepts of data privacy, its importance, threats, and mitigation over the development of mobile networks in multiple generations. 
    \item \textbf{Identify and distinguish personal and non-personal data in B5G/6G:} We explore how these two data types (personal / non-personal data) differ from each other and their unique characteristics.
    \item \textbf{Provide privacy goals in B5G/6G for personal and non-personal data:} Conceptualize a set of goals to be reached in each data type based on gaps identified in existing works.
    \item \textbf{Identify privacy issues for personal and non-personal data in B5G/6G:} Discuss the potential privacy vulnerabilities, threats, and attacks possible in B5G/6G networks.
    \item \textbf{Discuss potential solutions to solve the issues:} We provide a comprehensive discussion on possible approaches that can be used as mitigation mechanisms for privacy issues.
    \item \textbf{Summarise lessons learned and future directions for data privacy in B5G/6G:} We present our key outcomes and possible approaches that can ensure the privacy of data in 6G.
\end{itemize}

\begin{table*}[ht]
\scriptsize
\caption{Summary of important surveys on 6G privacy of personal and non-personal data\vspace{-4mm}}
\label{tab:surveys}
\renewcommand{\arraystretch}{1}
  \begin{tabular}{| p{0.2cm}|m{0.33cm}|m{0.33cm}|m{0.33cm}|m{0.33cm}|m{0.33cm}|m{0.33cm}|p{8.7cm}|}
  \hline
      \rowcolor[HTML]{9FC5E8} 
    	\multicolumn{1}{|c|}{\textbf{Ref.}} 
         & \multicolumn{1}{|c|}{\textbf{{\rotatebox[origin=c]{90}{6G Network}}}}
         &\multicolumn{1}{|c|}{ \textbf{{\rotatebox[origin=c]{90}{Taxonomy, Privacy Goals}}}}
         &\multicolumn{1}{|c|}{\textbf{{\rotatebox[origin=c]{90}{6G Personal Data}}}}
         &\multicolumn{1}{|c|}{\textbf{{\rotatebox[origin=c]{90}{6G Non-personal Data}}}}
         &\multicolumn{1}{|c|}{\textbf{{\rotatebox[origin=c]{90}{Privacy Solutions}}}}
         &\multicolumn{1}{|c|}{\textbf{{\rotatebox[origin=c]{90}{Future Directions}}}}
         &\multicolumn{1}{|c|}{\textbf{Remarks and Limitations}}
         \\ [10ex]
    \hline
    \hline
    \multicolumn{1}{|c|}{\cite{sandeepa2022survey}} & 
        \cellcolor{green!15} H & \cellcolor{yellow!20} M & \cellcolor{red!15} L & \cellcolor{red!15} L &
        \cellcolor{yellow!20} M & \cellcolor{yellow!20} M &
        Discusses on the privacy of B5G/6G networks using a layered architecture of 6G vision. Does not focus on data aspects but rather a general privacy discussion on 6G. Does not discuss non-personal data or differentiate personal data from non-personal data.\\ [2ex]
    \hline
    \multicolumn{1}{|c|}{\cite{nguyen2021security}} & 
        \cellcolor{green!15} H & \cellcolor{yellow!20} M & \cellcolor{red!15} L & \cellcolor{red!15} L &
        \cellcolor{yellow!20} M & \cellcolor{yellow!20} M &
        Overview of vulnerabilities, technological difficulties, security, and privacy solutions for 6G. Limited discussion on privacy compared with security aspects. No sufficient discussion on privacy aspects based on data types.\\ [2ex]
    \hline
    \multicolumn{1}{|c|}{\cite{wang2020security}} & 
        \cellcolor{green!15} H & \cellcolor{red!15} L & \cellcolor{red!15} L & \cellcolor{red!15} L &
        \cellcolor{red!15} L & \cellcolor{yellow!20} M &
       Security and privacy issues in 6G, and upcoming research challenges. Used security and privacy terms in combination that may be ambiguous to identify privacy issues separately. No significant discussion related to personal and non-personal data.\\ [2ex]
    \hline
    \multicolumn{1}{|c|}{\cite{sun2020machine}} & 
        \cellcolor{yellow!20} M & \cellcolor{red!15} L & 
        \cellcolor{red!15} L &
        \cellcolor{red!15} L &
        \cellcolor{yellow!20} M & 
        \cellcolor{yellow!20} M &
        Presents privacy of machine learning related to 6G. Privacy leakages from ML models and open ML issues are discussed. However, it does not specifically discuss personal or non-personal data aspects with ML or 6G.\\ [2ex]
    \hline
    \multicolumn{1}{|c|}{\cite{zhao2020comprehensive}} & 
        \cellcolor{green!15} H & \cellcolor{red!15} L & \cellcolor{yellow!20} M & \cellcolor{red!15} L &
        \cellcolor{red!15} L & \cellcolor{yellow!20} M &
        6G networks technologies, 
potential security, and privacy  
issues from the technologies. No specific focus on the privacy of personal or non-personal data.\\ [2ex]
    \hline
    \multicolumn{1}{|c|}{\cite{shahraki2021comprehensive}} & 
        \cellcolor{green!15} H & \cellcolor{red!15} L & \cellcolor{yellow!20} M &
        \cellcolor{red!15} L &
        \cellcolor{yellow!20} M & 
        \cellcolor{red!15} L &
        6G requirements and trends, 
revolutionary technologies, 
6G challenges and future research directions. The main discussion is generic and does not present much information on privacy in 6G or data.\\ [2ex]
    \hline
    \multicolumn{1}{|c|}{\cite{de2021survey}} & 
        \cellcolor{green!15} H & \cellcolor{red!15} L & 
        \cellcolor{yellow!20} M &
        \cellcolor{red!15} L & 
        \cellcolor{yellow!20} M &
        \cellcolor{yellow!20} M &
        Current developments of 6G, 
        Limitations of existing 5G mobile networks,
        new technology enablers for 6G,
        and standardization approaches. No significant discussion on goals and issues relevant to 6G personal and non-personal data privacy.\\ [2ex]
    \hline
    \multicolumn{1}{|c|}{\cite{finck2020they}} & 
        \cellcolor{red!15} L & \cellcolor{red!15} L & \cellcolor{yellow!20} M & \cellcolor{yellow!20} M &
        \cellcolor{red!15} L & \cellcolor{red!15} L &
        Provides a discussion on personal and non-personal data differentiation, classification, and examples. However, the work is not related to 6G and rather a generic description of data privacy.\\ [2ex]
    \hline
        \multicolumn{1}{|p{0.2cm}}{\textbf{This paper}} & \cellcolor{green!15} \textbf{H} & \cellcolor{green!15} \textbf{H} & \cellcolor{green!15} \textbf{H} & \cellcolor{green!15} \textbf{H} & \cellcolor{green!15} \textbf{H} & \cellcolor{green!15} \textbf{H} & \textbf{A comprehensive survey on B5G/6G personal and non-personal privacy taxonomy and goals, possible challenges and issues, potential solutions for the issues, and future research directions.}\\[2ex]
    \hline
  \end{tabular}
 
\begin{flushleft}
\begin{center}
\begin{tikzpicture}

\node (rect) at (0,0) [draw,thick,minimum width=1cm,minimum height=0.2cm, fill= red!15, label={[align=left]right:Low Coverage: The paper did not consider this area or only very briefly discussed it through mentioning it in passing}] {L};
\node (rect) at (0,0.4) [draw,thick,minimum width=1cm,minimum height=0.2cm, fill= yellow!20, label={[align=left]right:Medium Coverage: The paper partially considers this area (leaves out vital aspects or discusses it  without a specific focus on it)} ] {M};
\node (rect) at (0,0.8) [draw,thick,minimum width=1cm,minimum height=0.2cm, fill= green!15, label={[align=left]right:High Coverage: The paper considers this area in reasonable or high detail}] {H};
\end{tikzpicture}
\end{center}

\end{flushleft}
\vspace{-3mm}  
\end{table*}

\vspace{-1mm}
\subsection{Outline}
\begin{figure}[ht]
    \centering
    \vspace{-2mm}
    \includegraphics[width=0.7\textwidth]{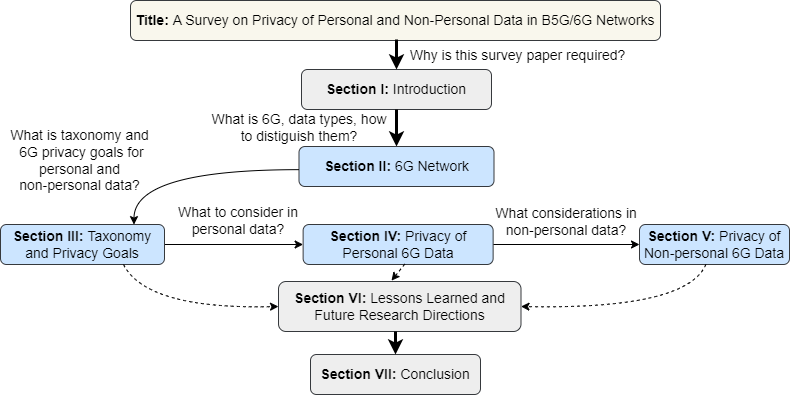} 
    \vspace{-3mm}
    \caption{Outline of the survey\vspace{-3mm}}
    \label{Figure_outline}
\end{figure}

The remaining sections are arranged as follows: Section~\ref{sec:6GNetwork} outlines the evolution of networks to 6G, 6G requirements, personal and non-personal data, and its importance to B5G/6G networks. Section~\ref{sec:TM} discusses the privacy taxonomy with goals to be achieved related to different data types. Section~\ref{sec:personal} provides details on issues, challenges, and potential solutions related to personal data. The issues related to non-personal data and possible methods to be used as solutions are provided in Section~\ref{sec:non-personal}. The outcomes and possible future directions are provided in Section~\ref{sec:lesson}. Finally, Section~\ref{conclusion} concludes the paper. Figure~\ref{Figure_outline} provides an outline of the arrangement of our survey.
\vspace{-3mm}
\section{Background}\label{sec:6GNetwork}

The progression of wireless networks has come a long way from several decades to the present. It has passed multiple generations, where each new generation sees significant progress than the previous generation's coverage, reliability, speed aspects, etc. More importantly, it can be seen that each new generation has given better consideration on data privacy than the previous, with advances in techniques used to ensure privacy. 
\vspace{-3mm}
\subsection{Evolution of Mobile Networks}
From the beginning of 1G in the 1980s to the current 5G era, data privacy has played a key consideration in mobile communications and networking in each generation. The 1G telecommunication networks were analog systems that used radio signals of frequency of 150 MHz, with only voice transmission~\cite{salih2020evolution}. There were many privacy issues in 1G, such as unencrypted communications with no guarantees of security or privacy. It led to problems such as masquerading, eavesdropping, and illegal access to data~\cite{wang2020security}. 

The 2G networks were introduced in 1991, with digital signals that facilitated clear voice calls, text, and multimedia messages via mobile phones~\cite{salih2020evolution}. There is a significant data privacy improvement in 2G networks, with digital modulation secured with encryption and temporary identifiers for anonymity. Yet it also had several security loopholes, as only one-way authentication of a user is allowed, where the user cannot authenticate the network, paving the way for illegal devices such as base stations to steal user data~\cite{wang2020security}.   

In 2000, the 3G UMTS was introduced, and they are mainly focused on delivering increased bit rates to satisfy the expected demand from users on data-intensive services, such as video telephony and downloads~\cite{mannweiler2020evolution}. Most of the privacy issues in 2G were mitigated in 3G. The 3G networks use Temporary Mobile Subscriber Identity (TMSI) instead of International Mobile Subscriber Identity (IMSI)~\cite{arapinis2012new}. IMSI is a permanent and unique identifier that can track the identity of a mobile subscriber. With TMSI, the privacy of these mobile subscribers is preserved as they are temporary and encrypted using a session key established in 3G Authentication Key Agreement process (AKA). However, 3G also came with several privacy issues, including caching of IMSI and 3G/GSM-interoperability privacy leakages~\cite{arapinis2012new}. 

The Fourth Generation of mobile networks was introduced in 2011, with the LTE standard to fulfill the demands of mobile internet traffic. It is designed to support only packet-switched services, in contrast to the previous generations that used circuit switching~\cite{mannweiler2020evolution}. The 4G has strengthened signaling protocols through authentication and ciphering more than the previous generations~\cite{shaik2015practical}. However, 4G also has data privacy vulnerabilities. They include fine-grained location leaks, heterogeneous wireless networks causing privacy issues with different data types, and incompatibilities in the encryption of 3G networks with 4G~\cite{shaik2015practical,gobjuka20094g}. 

The latest addition to the mobile networks is the fifth generation, where the 5G New Radio or 5G NR version of this generation was released in 2018~\cite{mannweiler2020evolution}. From the survey in~\cite{khan2020survey}, the first phase of the 5G radio interface was focused on addressing privacy vulnerabilities and threats. Examples include IMSI-catching and IMSI-probing attacks, unauthenticated International Mobile Equipment Identity (IMEI) requests,  and Globally Unique Temporary ID (GUTI) tracking. The 5G radio interface has various new features that enhance subscription privacy, such as false base station detection frameworks, secure radio re-directions, and frequent GUTI refreshment~\cite{khan2020survey}. However, there are various privacy concerns in 5G, within the radio interface and beyond, to the application levels. The works in~\cite{liyanage20185g,khan2019survey} comprehensively discuss the issues in 5G and applications privacy, which will be relevant for upcoming networks beyond 5G. Those works include ambiguity in data ownership, loss of governance/control, IoT privacy, issues in trans-border information flows, use of third parties, and having different objectives for trust. 

\vspace{-3mm}
\subsection{Transition from 5G to 6G}

It is anticipated in the 2030s that society will be highly intelligence-inspired, digitalized, and globally data-driven, with near-instant full wireless connectivity~\cite{latva2018radio}. The features provided by B5G/6G will undoubtedly open gateways to a wide range of associated applications that do not exist today. Most of these can be anticipated to be associated with providing ultra-realistic beyond current sensory inputs such as visual or auditory~\cite{liu2020vision} but extending to other senses as well. Also, it could be assumed that billions of devices will be connected, providing much more capabilities by fully digitalizing society and the underlying infrastructure, with use cases such as smart government and smart transport. The coverage will allow providing universal public services, despite the location boundaries due to B5G/6G ubiquitous coverage~\cite{liu2020vision}. 

The current 5G networks tend to have characteristics of migrating applications to the cloud, software-driven networking, virtualization, and slicing, but 6G will have all these features with an extra layer of intelligence added~\cite{zhang20196g}. It can be expected a substantial increase in automation by 2030, and the creation of distributed intelligence to support automation will be commonplace in these future networks~\cite {han2018network}. The future generation networks will also come with many smart applications that use this decentralized, distributed intelligence. Their application scenarios could be categorized into three forms: 1) intelligent life and intelligent interaction, 2) intelligent production, which supports scenarios such as smart agriculture, and intelligent industry-related applications, and 3) super transportation with full self-navigation in an intelligent society~\cite{liu2020vision}. 
All these applications will transmit massive amounts of big data every day through B5G/6G networks. This data can be related to an individual, or it can be non-personal and machine-generated. However, it is clear that in the near future, we will experience an explosion of data collected, processed, and analysed in almost every aspect of our life through the convenience of these networks. 

\vspace{-3mm}
\subsection{6G Requirements}
The data transmission in 6G networks is expected to be better than previous generations in almost every aspect, including data delivery, latency, accuracy, quality of service, and efficiency. Furthermore, the future networks will be capable of handling ultra-dense networks 10-100 times that of current 5G, which consists of billions or trillions of interconnected IoT, sensors, mobile devices, mixed reality devices, etc. They will generate massive amounts of data, capturing multi-dimensional sensor data. 6G is expected to deliver data rates 100 to 1000 times faster than 5G with up to terabit-per-second speeds~\cite{yang20196g}  to cope with the speed challenge. The work in~\cite{liu2020vision} shows that 6G should consist of a drastically improved user-experienced data rate, which should also be 100 to 1000 times higher when compared with 100 Mbps downlink user-experienced rate in 5G. Also, the control plane latency should be significantly lower in 6G, making applications that require rapid responsiveness possible when remotely operated, which is still a far-fetched idea with a latency of 10 ms in 5G. Therefore, B5G/6G networks will primarily fulfill the targets of ultra-high speeds and complex connectivity with very low latency.

\vspace{-3mm}
\subsection{Personal / Non-personal Data in 6G}
\label{sec:data_types}
The Big data collected within B5G/6G infrastructure can be broken down into numerous categories. In the privacy scope, we divide them into personal and non-personal data, depending on the specific properties of data generation. The following subsections describe the definitions, properties, differentiating between the data types, and the summary of key points.
\vspace{-2mm}
\subsubsection{Personal Data in 6G}
According to GPDR definition Art. 4~\cite{gdprArt4}, there are two categories of data considering the linkage with the data source: personal data and non-personal data. Personal data is the data directly relating to an individual, which can be used to identify the natural person (data subject). It should be noted natural person here differentiates from a juridical person. Natural person refers to a human person~\cite{adriano2015natural}, which distinguishes it from corporate entities/ collective person/legal entity, which is regarded as non-human. 

Personal data consists of direct or indirect properties that can link to a natural person. This could be directly or indirectly identifiable through a reference such as a name, an identification number, location data, an online identifier, or to one or more factors specific to the physical, physiological, genetic, mental, economic, cultural, or social identity of that natural person~\cite{gdprArt4}. Depending on the significance of privacy, or individual preference, such data can be sensitive, private, or not. Thus, personal data may not always be regarded as private and may be shared publicly. Otherwise, they are regarded as sensitive personal data, which should be private and protected. Examples of sensitive private data may include physical properties such as biometrics and genetic data, or they may be conceptual, such as personal beliefs and political opinions. A person may have control over the level of exposure and sharing of their personal data, which makes them either private, selectively private, or public. 

In the context of B5G/6G networks, there is a high likelihood of introducing new types of personal data and getting adopted by a majority of end users. They may include technologies that will be facilitated by the 6G network capabilities and extensive data rates. Examples include Augmented Reality (AR)/Virtual Reality (VR) and metaverse-based new data types, which include data such as digital avatars of a real person, live perceptions, and user actions in real-time~\cite{mystakidis2022metaverse}. These data may also link with personal emotional states, which can often be considered private and sensitive data.


\vspace{-2mm}
\subsubsection{Non-Personal Data in 6G}
\begin{figure}[htb]
    \centering
    \includegraphics[width=0.9\linewidth]{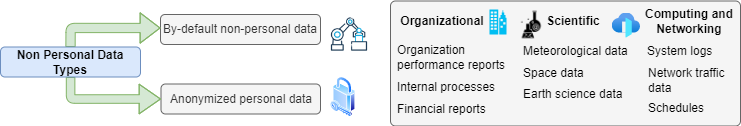} 
    \vspace{-3mm}
    \caption{Non-Personal data types and examples of non-personal data in domains of organizational, scientific, computing, and networking\vspace{-4mm}}
    \label{Figure_np_data}
\end{figure}

According to~\cite{finck2020they}, non-personal data is given as a type of data that is never related to an identified or identifiable natural person. There are also two categories~\cite{finck2020they} considering non-personal data: 1) anonymized data, which once was personal data where the linkage to the person is removed, 2) by-default non-personal data, which does not relate to a person; thus anonymization may not be needed. Figure~\ref{Figure_np_data} shows non-personal data categories and examples in different fields.

\vspace{-3mm}

According to the Recital 26 GDPR~\cite{gdprArt4}, anonymous data is not considered for principles of data protection. Therefore, it falls outside the scope of European data protection law~\cite{finck2020they}. However, in the EU, the Free Flow of non-personal Data (FFD) Regulation framework provides regulation approaches and guidelines for this type of data. The regulation outlined in~\cite{nonpersonaleu} mentions the emerging significance of non-personal data with the industrialization and aggregation of big datasets. FFD also provides some examples and use cases of non-personal data, such as aggregate and anonymized datasets used for big data analytics, data on precision farming, and data on maintenance needs for industrial machines. They can be applications in B5G/6G as well.

The initial notion of non-personal data can be found with different terms describing such as machine-generated data, industrial data, data created by computer processes, IoT data, and sensor data~\cite{wendehorst2017elephants}. Clearly, such generic terms are not completely applicable for non-personal data since they may contain data that are directly or indirectly related to an individual~\cite{wendehorst2017elephants}. Therefore, a general characteristic of non-personal data will not have a direct or indirect relationship with a natural person and will pose a minimum likelihood of having a linkage or singling out of an individual. Thus, non-personal data can provide details of a certain non-personal entity that can include a value for data owners on that entity. They are identified as one of the key economic assets, despite issues of the future possibilities of linkage or consent upon ownership~\cite{janevcek2018ownership}. 
With future B5G/6G networks, there will be a surge of emerging application domains such as hyper-intelligent IoT, smart cities, industry 5.0, and autonomous vehicles~\cite{de2021survey}. All these applications will produce machine-generated, anonymized, and unlinked data that can be regarded as non-personal. Further, they will be generated in huge quantities at a much more rapid phase than current data generation speeds. Real-time tracking, aggregation, processing, secure storage, and sharing of such non-personal data will be challenging issues that need further discussion.

\vspace{-2mm}
\subsubsection{Borderline between personal data and non-personal data:}
\begin{figure}[ht]
    \includegraphics[width=0.8\linewidth]{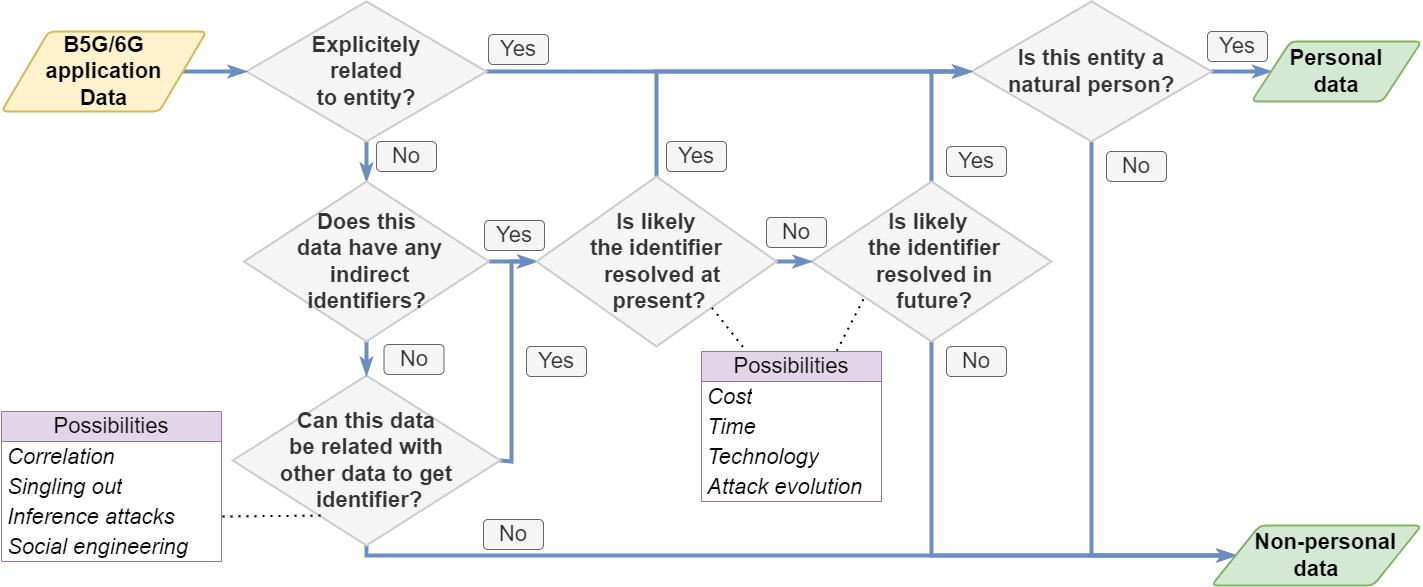} 
    \vspace{-3mm}
    \caption{Assessing data to classify and differentiate personal and non-personal data. The classification may consist of multiple factors to be taken into account when direct personal information is not available.}
    \label{classification_personal_nonpersonal}
    \vspace{-3mm}
\end{figure}

There may not be a clear border between personal data and non-personal data due to the future possibility of the resolution of anonymized data in 6G networks via novel modes of privacy attacks. Furthermore, non-human data may still consist of certain attributes that can link with individuals due to the involvement of end users or operators with the machine-generated data. However, the work in~\cite{finck2020they} provides a test for distinguishing personal data from non-personal data devised by Recital 26 GDPR. Figure~\ref{classification_personal_nonpersonal} provides an overview of the test.

\vspace{-2mm}

The factors considered here are the possibility of the resolution of a natural person with the availability of additional knowledge, the correlation of existing data with a person, and the likelihood of resolution. We added the present and future likelihood since factors such as cost and time might get significantly reduced for the resolution of anonymized data with technology improvements. The rollout of 6G is expected in the early 2030s, which will involve many possibilities of introducing new types of data analytics, inference, reconstruction attacks, and a significant technological advancement in computation capabilities. Due to this dynamic nature, it would be difficult to assure a life-long "non-personality" of data~\cite{somaini2020regulating}. Thus, it is of paramount importance to evaluate the possibilities of threats, issues, and potential solutions for non-personal and personal data protection, to preserve the privacy of individuals using the B5G/6G services.

\vspace{-3mm}
\subsection{Role of Personal Data and its Importance for B5G/6G} 

\begin{figure}[ht]
    \includegraphics[width=.9\linewidth]{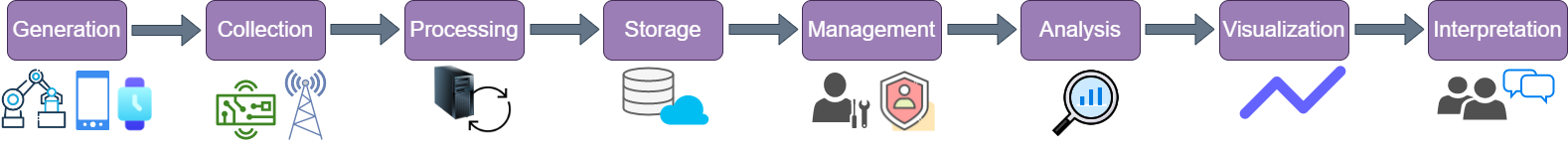} 
    \vspace{-4mm}
    \caption{The data life cycle involves multiple steps, from generation to the interpretation phase. Multiple actions and actors are involved in this process for B5G/6G.} 
    \label{data_life_cycle}
    \vspace{-4mm}
\end{figure}

Raw data from a data source involves a life cycle with multiple intermediate steps~\cite{wing2019data}. Figure~\ref{data_life_cycle} represents these steps in the process. The FFD framework~\cite{nonpersonaleu} also provides different activities in a data value chain: data creation and collection, aggregation and organization, processing, analysis, marketing, distribution, use, and re-use of data.
Personal data will include identity of the natural person in the processing and storage phases. Thus, the proceeding analysis, visualization, and interpretation phases may provide insights into individual data generators, which may be used for commercial purposes. The B5G/6G will involve in each stage of this life cycle since data should be moved from the source to a remote destination in large quantities. Personal identifiers may provide the possibility to personalize services offered by B5G/6G end users. For example, applications such as dynamic spectrum allocation via cognitive radio will increase the efficiency of spectrum  by allocating spectrum holes of primary users to secondary users~\cite{ahmad20205g} in B5G/6G. These applications may require the personal data of users such as geolocation~\cite{grissa2019location} to determine to whom the spectrum should be properly allocated. 

Another dominant area predicted in B5G/6G is the native support for AI by design~\cite{wu2021toward} and the paradigm shift of cloud AI towards network/edge AI~\cite{an20216g,xiao2020toward}. For this transition to happen, it is clear that personal data will be heavily used for AI algorithms to train their model. Privacy-preserved new learning techniques such as Federated Learning (FL) will allow AI model training with freshly generated personal data within the user devices or in the edge~\cite{lim2020federated}. Thus, timely updated AI models will inherently make the management, orchestration, and monitoring of the 6G network and applications. Therefore, personal data perform a highly impactful role in B5G/6G networks to deliver guarantees towards a fully autonomous network infrastructure. 

\vspace{-3mm}
\subsection{Role of Non-Personal Data and its Importance for B5G/6G}

A tremendous amount of non-personal data is also expected to be generated in the 2030s era, with about 500 billion new IoT devices introduced~\cite{zikria2021next} to the network. Non-personal data may also include many output logs and statuses generated by automated processes in massive amounts of 6G operations since these operations and orchestrations are driven with Zero-touch network and Service Management (ZSM)~\cite{liyanage2022survey} in B5G/6G networks. 

The non-personal data in 6G will also be anonymized and perturbed data which were originally personal data. These perturbations may be done in the initial phases of the data life cycle in Figure~\ref{data_life_cycle}, such as the generation, collection, and processing stages, to remove the linkage with individuals. The likelihood of recovering personal information from anonymized data will be reduced with the advancement of perturbation and anonymization techniques in the B5G/6G era. Many privacy-preserving techniques such as homomorphic encryption, differential privacy techniques, and secret sharing~\cite{sun2020machine,kakkar2020survey}  can be expected to continuously improve and will act as enablers for "non-personal nature" in data. Such non-personal data can also be used for more generic applications like crowd-sensing, traffic monitoring, emergency situation detection, AI model training applications for anomaly detection, etc. With up-to-date anonymized or perturbed data from 6G networks shared, it will be possible to improve the predictability of AI models with the latest data. Therefore, non-personal data directly play a massive role in B5G/6G network applications, especially concerning the capability of sharing data among multiple parties without compromising individual privacy.

\vspace{-2mm}
\section{6G Privacy Taxonomy and Goals} \label{sec:TM}
This section outlines different taxonomies of privacy that can be incorporated into B5G/6G. Based on the taxonomy, we formulate a set of privacy goals, which we identify as essential requirements to be specifically achieved in B5G/6G networks. We divide these goals based on their relevancy and significance for personal and non-personal data. Both goals relevant to personal and non-personal data are classified as hybrid privacy goals.

\vspace{-3mm}
\subsection{6G Privacy  Taxonomy}

\begin{figure}[ht]
    \vspace{-3mm}
    \centering
    \includegraphics[width=\linewidth]{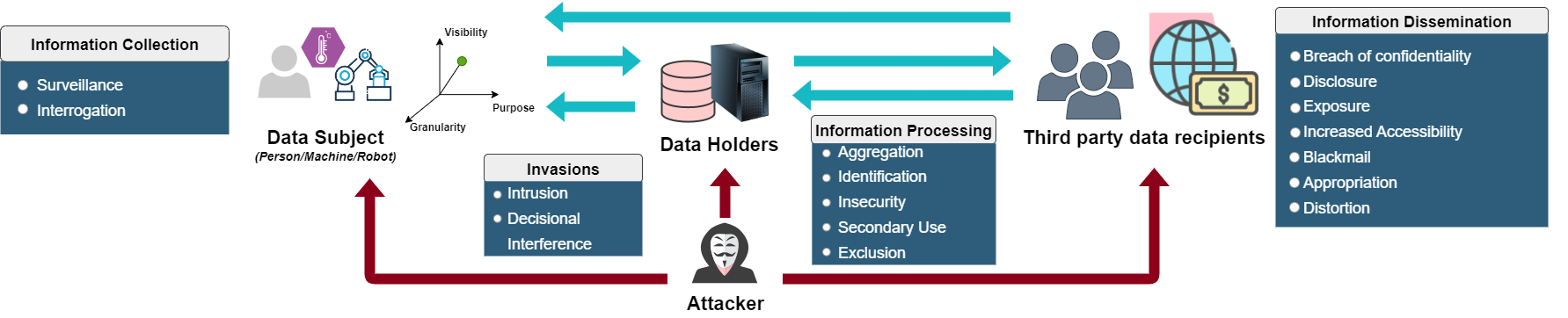} 
    \caption{The privacy of data by different actions performed in B5G/6G. Privacy settings are defined by the data generators and shared by data holders. Attackers may try to exploit the weaknesses in the process to recover private information.}
    \label{taxonomy_privacy}
    \vspace{-3mm}
\end{figure}

Privacy has a long development history, and its definitions change depending on the era, society, and the individual~\cite{lukacs2016privacy}. With the enormous amount of data collected on individuals, privacy is mainly focused on data privacy. The GDPR~\cite{gdpr_eu_2019} states that data privacy gives users the ability to choose who can use their data and for what purposes. The work in~\cite{barker2009data} describes three dimensions of contributors for data privacy: visibility, granularity, and purpose, and assigns different levels for each contributor. For example, visibility has different levels, such as from no access, only data owners, and third parties to the public. Similarly, the purpose can be defined as single, reuse same, reuse selected, and any use of the data. The granularity can be existential, partial, or specific data. According to the user preference, they can adjust the level of privacy for the data from these levels. The authors in~\cite{solove2005taxonomy} present a taxonomy of data privacy based on different actions that are done on data. An overview of the actions and their relationship is provided in Figure~\ref{taxonomy_privacy}. They are as follows:

\begin{itemize}
    \item \textbf{Information collection} - The collection stage can occur via surveillance and interrogation. Third-party services may do this in B5G/6G, and data from natural persons and machine-generated non-personal data will be collected. During this stage, the option for users to select their privacy preferences in~\cite{barker2009data} can be provided.

    \item \textbf{Information processing} - The collected data is processed by third-parties. For B5G/6G, AI-based services will be used for processing personal and non-personal data, which can reveal unintended sensitive information or linkage with other individuals. 

    \item \textbf{Information dissemination} - The privacy during sharing data is considered here. The dissemination may cause privacy issues such as a breach of confidentiality that was initially agreed upon among the parties, an increase of accessibility that was previously not intended, and distortion of original information~\cite{solove2005taxonomy}. This step may need to include proper access controls and privacy-preserving mechanisms.

    \item \textbf{Invasion} - An attacker or an eavesdropper may attempt to evade the privacy preservation mechanisms to recover original information, infer about specific attributes that are intended to be private, or gain control over the data. Such attackers may exploit the vulnerabilities in data generators, data transfers, storage, processing operations, privacy-preserving mechanisms, or data-sharing processes in B5G/6G services. The likelihood of such invasions is also a key consideration when defining non-personal data.

\end{itemize}

As it is clear that 5G and B5G networks can handle vast amounts of data, they may be subjected to privacy leakages due to the complexity of this extensive data and wide range of services. Potential issues in privacy can go undetected in these latest technologies and new services. For example, considering edge computation, with limited capability and lack of privacy protection, it could be possible for an adversary to exploit these issues on a large scale due to the more-than-ever inter-connectivity of the new generation networks. Therefore, we propose a set of privacy goals that need to be achieved to ensure these networks are trustworthy for general users. In~\cite{porambage2021roadmap}, authors provide open challenges for privacy for 6G. The work in~\cite{liyanage20185g} defines a set of objectives for privacy in 5G networks. By extending this information, we provide our privacy goals for B5G/6G as follows:

\vspace{-3mm}
\subsection{6G Personal Data related Privacy goals}
\subsubsection{Ensure the personal data privacy during data generation} 
The origin of personal data in B5G/6G is often connected with IoT sensor environments. The authors in~\cite{nguyen20216g} classify five domains of 6G IoT, namely healthcare IoTs, Vehicular IoTs and Autonomous Driving, Unmanned Aerial Vehicles, Satellite IoTs, and Industrial IoT. These domains may consist of possibilities of capturing personal data like health data, location, actions, and habits, which are often considered private and sensitive. When generating such data and transmitting it to nearby devices, a guarantee of protection for raw data is important. Since the IoT sensor environments have relatively weaker privacy mechanisms, they can be easy targets for privacy attacks~\cite{meneghello2019iot} via compromising the devices or eavesdropping on the communication. Therefore, to deliver personal data safely and privately, significant improvements in lightweight yet robust privacy mechanisms can be considered in the future. Consequently, we consider safeguarding this first step of the data generation and collection process as a critical goal to be reached for personal data in B5G/6G.
\vspace{-2mm}
\subsubsection{Privacy guarantees for edge networks and edge AI}
With edge computing, sensor-generated personal data can be transmitted to shorter distances with low energy and processed close to or at the network's edge. It improves the QoS and user experience.
However, trusting edge devices is a challenge that future networks will face. This is because untrusted third parties can connect their devices and services to the network~\cite{du2018big}. Therefore, we think having guarantees for edge networks is an important factor since there is an explosive increase in the number of interconnected edge devices with the 5G~\cite{liu2020toward}. It will continue in the upcoming generations. Edge AI is another crucial area rising with the edge networks, which can be used to offload decision-making from the cloud to the edge devices. This will reduce the cloud servers' load and traffic by significantly filtering unnecessary data. It is also used as a data privacy preservation technique through methods such as FL, which will be discussed further in Section~\ref{sec:solutions}. Therefore, with the rise of this edge intelligence, the guarantees for edge privacy should also consider safeguarding the edge AI's privacy.
\vspace{-2mm}
\subsubsection{Clarify the responsibility and accountability for entities processing personal information}
The transparency of data explainability is a primary requirement in B5G/6G networks, with the increased complexity of network connectivity. Also, due to the wide range of services and many intermediates, it will not be easy to understand who will be responsible for users' privacy. For instance, in the case of IoT, the work in~\cite{aleksandrova2021right} discusses the issues in IoT relevant to data deletion. It also shows that an Ambient Assisted Living (AAL) system relies on various sensors and smart home appliances from different manufacturers with different communication concepts. Then, a question arises about who is the data controller. This may raise the question of who will be responsible for selecting the data controller and updating the contents. Also, an explanation on what grounds the decision is taken regarding this may be required. This is an important privacy goal to be considered in B5G/6G networks.
\vspace{-2mm}
\subsubsection{Guarantee of erasure and rectification of personal data}
The right to erasure or the right to be forgotten is also an emerging privacy requirement that is considered in legal systems~\cite{connolly2021right,cabral2020forgetful}. A person should be able to request to rectify incorrect or incomplete personal data or to erase them from the digital records~\cite{liyanage20185g}. The goal is to ensure this erasure. As many connected parties are involved in B5G/6G future networks, it is required to update all the parties on the update or erasure of this user information.
\vspace{-2mm}
\subsubsection{Achieving personal data privacy in AI-based training and processing operations}
The B5G/6G services and network components are expected to work under full automation from AI. For this, powerful AI models are required in the future that may need to be trained with millions and billions of actual examples of personal data like user behaviour, habits in using the network, preferences, network traffic data, etc. The AI frameworks may also share this model among third parties. For example, FL trains an ML model within the end-user device and then forward this local model to a third-party aggregator. The aggregator joins the model to create a global model and share it with other local devices. Thus, local model parameters trained with personal data are shared with others, which can lead to privacy issues such as membership and attribute inference attacks~\cite{mothukuri2021survey}. Furthermore, AI models can identify patterns from data that may not be intended to be exposed by individuals. Examples include AI-based cyberattacks~\cite{yamin2021weaponized} and explainable AI (XAI)-based model extraction attacks~\cite{yan2022towards}. There are many proposed mitigation mechanisms in the existing literature, such as perturbation~\cite{rahman2018membership}, regularization~\cite{yamin2021weaponized}, and masking confidence scores~\cite{jia2019memguard} which will be discussed in Section \ref{sec:solutions}. The goal in B5G/6G should be to maintain the model utility to perform the operations with higher accuracy while preserving the privacy of the personal data that is being processed.
\vspace{-2mm}
\subsubsection{Getting explanations of AI actions for privacy requirements}
The users have the right to question decisions made by AI that handles their personal data. Therefore, AI used in B5G/6G network operations should be explainable, and responsible entities should explain how their AI made that decision and the possible assumptions. However, many of today's machine learning algorithms put themselves in a conventional black box view. Therefore, AI explainability can be considered to be one of the most important goals in terms of privacy requirements.
\vspace{-2mm}
\subsubsection{Balance the interests in personal data privacy protection in global context}
There are already existing regional-level privacy frameworks for personal data, such as EU GDPR. However, for achieving a unified standardization, it may be an ambitious goal to make a global level understanding of the importance of privacy protection, satisfying the interest in fostering privacy services globally. While methods such as human rights mechanisms may help, they are insufficient on their own, and more robust protective standards for cyberspace activities should be developed~\cite{rojszczak2020does}. 
However, it is also essential to maintain the balance between privacy requirements and industry when imposing regulations in this case. The studies in~\cite{peukert2020european} show that websites have substantially reduced their interactions with web technology vendors after GDPR became effective. Also, they mention that many firms undergo losses while major players significantly increase their market shares. This makes the market less diversified and raises the barrier to entry. 
It implies a substantial effect on industries and complications in handling organizations' data with third parties. Thus, we identify finding the balance between personal data privacy and the interests of B5G/6G services applicable globally as an important goal to be reached.
\vspace{-3mm}
\subsection{Non-Personal Data related 6G Privacy Goals}

\vspace{-1mm}
\subsubsection{Proper identification and measurable distinction for non-personal data} 

As explained in Section~\ref{sec:data_types} non-personal data is assumed to have a minimum likelihood of being linked with natural persons. However, it is unlikely to measure this likelihood with an absolute amount or a level at the present moment. As discussed in section~\ref{sec:data_types}, the possibility of recovering personal data from anonymized non-personal data depends on many factors, including the expected technological progress, time, cost, and available resources for an adversary~\cite{finck2020they}. Identifying these factors and their impact is essential when defining non-personal data in B5G/6G networks. Also, suppose a quantifiable and clear boundary between personal and non-personal data exists. In that case, it will be easier for data to be used and shared among third parties, as privacy guarantees for such data exist to a degree of confidence.
\vspace{-1mm}
\subsubsection{Evaluation of potential privacy leakages from non-personal data}
Non-personal data is emerging attention for the B5G/6G networks due to the rapid addition of machine-generated data from various sources such as industrial robots, network sensing, and scientific equipment. As discussed, another form of non-personal data is the anonymized data of individuals, which were used to be personal data~\cite{finck2020they}. The increasing amount of non-personal data will inherently create the requirement to ensure privacy. Though they are not directly related to individuals, there could still be potential threats to people, especially to industries or organizations that generate these data. They might reveal sensitive individuals' or confidential organizational details if a third party carefully analyses this data. Examples of such non-personal data leakages include re-identification attacks, where adversaries may attempt to identify anonymized individuals from public datasets used for commercial or research purposes~\cite{shamsi2018understanding}. Hence, we consider it an important goal that should be fulfilled for the next generation of wireless networks.
\vspace{-1mm}
\subsubsection{Achieving privacy protected AI driven automated network management operations} 
The B5G/6G networks should fulfill massive network traffic demands for billions of connected devices with better QoS. Network management functions automation is essential to make this a reality. Such an example of an approach is ZSM by ETSI, where the authors in~\cite{benzaid2020ai} provide an architecture driven by AI for complete automation of management operations. However, it is also essential to regularly monitor the functionality of this automation for any privacy leakages since adversaries may find vulnerabilities in the AI models and decisions. Also, it is possible to manipulate these automated services and make attacks to extract non-personal data such as network behaviour, service logs, operations, and other private details of the network activity. Therefore, privacy protection is imperative when automating network operations in B5G/6G networks. 
\vspace{-1mm}
\subsubsection{Standardization of privacy in technologies, and applications for non-personal data}
The technologies and applications used in B5G/6G networks evolve independently. However, when considering the real-world use of networks, we see they are interconnected. They may use their different set of protocols to communicate with each other. Therefore, we consider the proper standardization of security mechanisms as an important step to ensure privacy protection for non-personal data during interfacing and communicating among applications of different technologies. This could be taken into account, especially when interfacing with edge devices, as they are more vulnerable to privacy leakages. The current laws~\cite{greenleaf2019global,goldman2020introduction} like EU GDPR,  Personal Information Security Specification in China, General Data Privacy Law (GDPL) of Brazil, and California Consumer Privacy Act (CCPA) consider personal data privacy. Relative to this, non-personal data privacy regulations need to be improved further to satisfy future requirements of data flow among different entities.~\cite{liyanage20185g} also shows that harmonizing privacy services in the global context to promote a digital single market is an important objective for B5G/6G networks. 
\vspace{-1mm}
\subsection{6G Hybrid Privacy Goals}


\vspace{-1mm}
\subsubsection{Ensure privacy protection for big data systems}
\begin{figure*}[ht]
    \vspace{-2mm}
    \centering
    \includegraphics[width=0.7\linewidth]{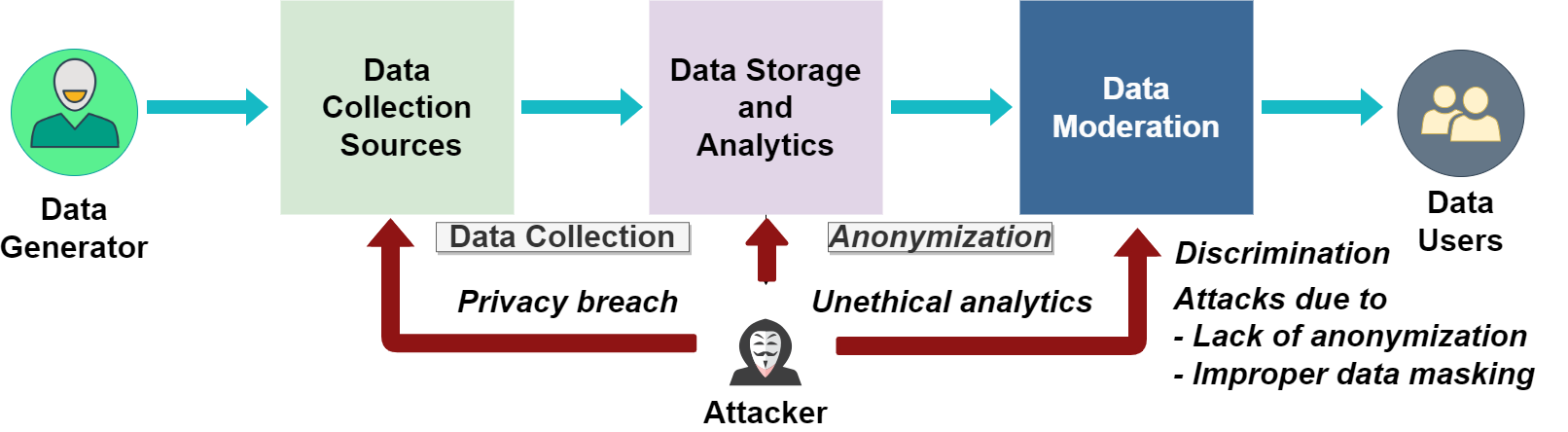}
    \caption{Big data system and privacy considerations on different stages of B5G/6G operation}\vspace{-4mm}
    \label{BgDataSys}
\end{figure*}

The amount of big data is ever-increasing with recent technology such as digital and cloud storage, IoT, and network. Both personal and non-personal data will be aggregated at huge volumes within a short period due to promising data rates and scalability features offered by B5G/6G. Therefore we can identify that B5G/6G networks contribute significantly to big data. Privacy should be considered in each stage; data generation, storage, and processing. Figure~\ref{BgDataSys} shows these stages and the possibility of attackers exploiting the process~\cite{yu2016big}. However, the work to enhance privacy in big data may not suffice the rate at which it is generated. There are existing techniques for big data privacy, including HybrEx, k-anonymity, T-closeness, and L-diversity~\cite{jain2016big}. Yet, more solutions for privacy enhancement should be given for B5G/6G networks to ensure user trust in them. 

\vspace{-1mm}
\vspace{-1mm}

\subsubsection{Achieving balance between privacy and performance of services}
The privacy-preserving algorithms may cause performance degradation with the implementation of methods such as encryption in the real world~\cite{zhang2016privacy}. This can be significant for devices with limited computation capacity, such as IoT, yet it is imperative to implement privacy preservation for such systems in B5G/6G. Achieving the right balance between privacy and performance is another goal that we can propose. One of the attempts to understand the balance in the context of IoT is in~\cite{dong2018quantifying}, where authors propose a two-step approach: first, get the trade-off between the quality of collected data and performance, and second, to understand how the nature of data affects adversary's ability to infer the user's private information. 
Furthermore, privacy-preserved AI-based decision-making may add further overhead to operations that must perform at terabit speeds. Therefore, achieving the balance between privacy and performance is important to maintain the QoS and data rates. 

\vspace{-1mm}
\subsubsection{Achieving proper utilization of interoperability and data portability}
Data portability is easily transferring data from one system to another without re-entering data. This can ease a consumer's life since they can avoid the redundant task of entering data in multiple applications in B5G/6G.

\begin{figure}[ht]
    \centering
     \begin{subfigure}[b]{0.3\textwidth}
         \centering
         \includegraphics[width=\textwidth]{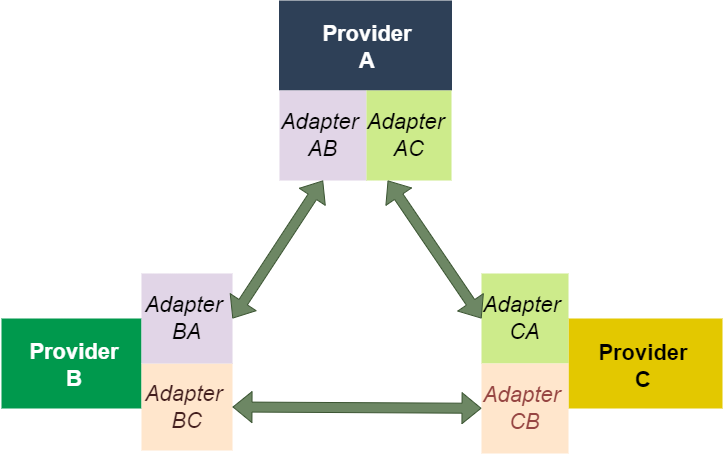}
         \caption{without data model}\vspace{-1mm}
     \end{subfigure}
      \hspace{5mm}
     \begin{subfigure}[b]{0.3\textwidth}
         \centering
         \includegraphics[width=\textwidth]{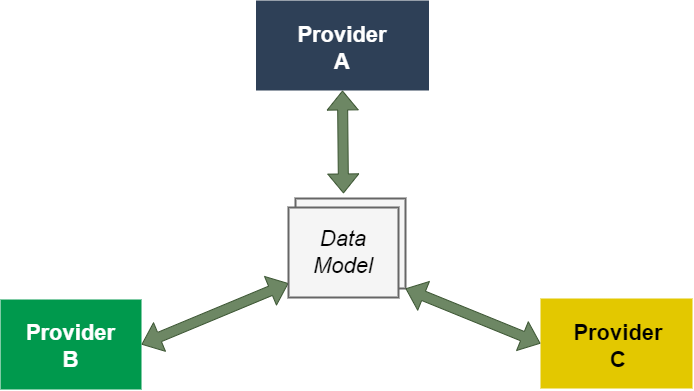}
         \caption{with data model}\vspace{-1mm}
     \end{subfigure}
     \hfill
    \caption{Overview of Data Transfer Project~\cite{datatransferproject}: (a) without data model, adapters have to be written for communication among providers, (b) with data model, only a single API is need to be maintained for communication with any provider}\vspace{-1mm}
    \label{dtp_architecture}
\end{figure}

We observe recent collaboration to enhance interoperability among companies, including Google, Microsoft, Twitter, and Apple. They have initiated a Data Transfer Project to allow individuals to transfer their data between online service providers~\cite{egan2019data}. It is an open-source project to extend data portability beyond a user's ability to download a copy of their data from their service provider, allowing the user to initiate a direct transfer of their data into and out of any participating provider. It shows that it is crucial to consider privacy principles of data minimization and transparency, with clear and concise data to transfer among such systems.
The project is implemented using adapters that convert various proprietary data formats to a small number of canonical forms or data models~\cite{datatransferproject}. This helps achieve data portability. It is easier to fulfill privacy requirements for a single data type in each organization than for different data types and multiple interfaces. Figure~\ref{dtp_architecture} shows an overview of the implementation of the project.
However, the right to data portability remains barely known among consumers and is only implemented in a fragmented manner~\cite{kuebler2021right}. Thus, several gaps may need to be fixed before reaching full data portability, which we consider a goal for both personal and non-personal data.

\vspace{-1mm}
\subsubsection{Quantifying privacy and privacy violations}
Privacy is difficult to define as it is a subjective concept and has different levels from person to person. Even the opinion on privacy that one may have may vary over time. Due to this nature, quantification of privacy can be difficult. However, based on the context, we may provide local standards and metrics related to different types of privacy. The process may need to incorporate other fields, such as psychological aspects. Therefore, quantifying privacy can be considered a challenging goal to be reached in B5G/6G to provide a sufficient guarantee of privacy for data.



\vspace{-1mm}
\section{PRIVACY OF PERSONAL 6G DATA}
\label{sec:personal}
In this section, we first discuss the privacy issues related to personal data in B5G/6G. Then, we focus on potential techniques that can act as solutions for the mentioned issues.
\vspace{-1mm}
\subsection{Issues and Challenges for Personal Data in B5G/6G} \label{sec:issues}

The personal data in 6G face several challenges that appear as barriers to the privacy goals mentioned. This section describes such issues in detail. In each given privacy issue, we provide an introduction to each, related works in the literature concerning the issue, and our opinion on the significance and the directions regarding it.
\vspace{-1mm}
\subsubsection{Privacy attacks on AI models and private data}
Different privacy attacks on AI models used in 6G can cause severe leakages in the privacy of individuals. Modern ML models inherently use big datasets to train and provide better predictions with large quantities of data. However, these models, therefore, can be regarded as a representative version of the original data used for training. Examples of privacy attacks on AI models include inference attacks such as membership inference, reconstruction attacks, and model extraction attacks~\cite{sandeepa2022survey}. The authors in~\cite{dwork2017exposed} also provide details on different attacks possible for private data, including re-identification, reconstruction, and tracing attacks. Such attacks can leak sensitive information about the B5G/6G users. The work~\cite{dwork2017exposed} also shows possible attacks on aggregate data, such as statistics on distributions like gender, disease, and genetic traits.

Another area is XAI, which is emerging as a tool to explain the black box AI models to understand the decisions made by AI. However, XAI can also be a "double-edged sword" regarding privacy of 6G~\cite{sun2020machine}. For example, the work in~\cite{yan2022towards} discusses the vulnerability of model extraction attacks with XAI. Furthermore, they mention XAI can reduce the difficulty in reconstruction attacks, and XAI can be used to improve the success rate of membership inference attacks.

\vspace{-1mm}
\subsubsection{IoT edge network and edge AI privacy attacks}
IoT is regarded as well established technology which is expected to expand immensely with future B5G/6G networks. However, there can be potential for privacy leaks due to its nature, like resource limitations, weaker privacy protocols, and easy, low-cost access. The work in~\cite{alferidah2020review} presents a taxonomy of IoT attacks based on different features such as device properties, adversary location, access level, attack strategy, protocol, etc. An example of a privacy attack launched in IoT is a data mining attack on smart meters~\cite{wang2021privacy}, where an adversary attempts to eavesdrop on the communication between the gateway and a target user. The work in~\cite{elhoseny2021security} describe possible attacks in Medical IoT (MIoT), where attackers can launch side-channel attacks to eavesdrop on the communication channel, sensor tracking to obtain location data of patients, and unauthorized access to obtain sensitive patient data. 

\begin{figure}[ht]
    \centering
    \includegraphics[width=0.7\linewidth]{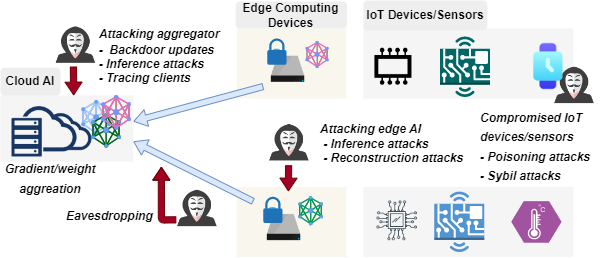} 
    \vspace{-3mm}
    \caption{Privacy attacks on B5G/6G FL edge AI system}
    \vspace{-6mm}
    \label{edge_privacy}
\end{figure}

FL is used as an emerging B5G/6G edge AI solution for resource-constrained devices. However, FL can undergo different types of attacks. One type of such attack is sybil attacks~\cite{alam2022federated}, where an attacker attempts to imitate many end devices without disrupting the FL protocol. This makes it hard to identify the attack due to the nature of FL. Other possible attacks include attacks on the training phase, such as poisoning, and attacks in the inference phase~\cite{alam2022federated}, such as membership inference. 
Figure~\ref{edge_privacy} provides an overview of the attacks on FL edge AI systems.
\vspace{-1mm}
\subsubsection{Privacy limitations in cloud computing and storage environments}

In the era of B5G/6G, vast collections of data will be read from many consumer products such as mobiles, computers, and sensor data in IoT devices. This makes it difficult to maintain in-house storage for many organizations. Therefore, they will usually be outsourced to third-party cloud storage providers who maintain the data in a distributed fashion. 
However, there are some privacy issues relative to data storage in such environments. 

The work in~\cite{ghorbel2017privacy} categorizes privacy issues in the cloud into four areas by considering privacy vulnerabilities: lack of user control, dynamic nature of the cloud, privacy compliance, and accountability. This work mentions the issues of data loss and leakage, illegitimate data handling, illegitimate data dissemination, and unauthorized secondary usage. Data duplication and retention occur due to the dynamic nature of the cloud. Authors in~\cite{ghorbel2017privacy} also mention that, in reality, data can be scattered in many geographical locations, and data owners may not have a clear idea of who can access it, who keeps it, and what happens when they request data deletion. A survey on sensitive data~\cite{domingo2019privacy} mentions two scenarios that affect privacy for the cloud: the large volume of sensitive data collected and the upgrades from data protection laws against outsourcing the processing of unprotected sensitive data.
The work in~\cite{ramirez2014data} shows that data brokers could use behaviour patterns to extract sensitive information from a user, which may expose sensitive details such as chronic diseases, religious interests, political affiliation, investment habits, etc. Privacy-preserving access control schemes can bring high complexity, which limits cloud services' scalability and flexibility~\cite{sun2019privacy}. The work in~\cite{zhang2020attribute} shows global properties of datasets (e.g., the average income of individuals in a dataset rather than an individual's income), which can be sensitive and confidential, even though the individual data points are anonymized. These attributes may easily be revealed through data analysis. The work~\cite{zhang2020attribute} further shows that revealing some global properties may leak trade secrets, intellectual property, and other information related to the data owner. This could happen even when techniques such as differential privacy (which is discussed in Section~\ref{sec:solutions}) are applied to individual data. 
Therefore, cloud computing still faces numerous privacy challenges to be addressed when adopting the new B5G/6G networks, though it is a reasonably matured technology. 
\vspace{-5mm}
\subsubsection{Cost on privacy enhancements}
\begin{figure}[ht]
    \centering
    \includegraphics[width=0.6\linewidth]{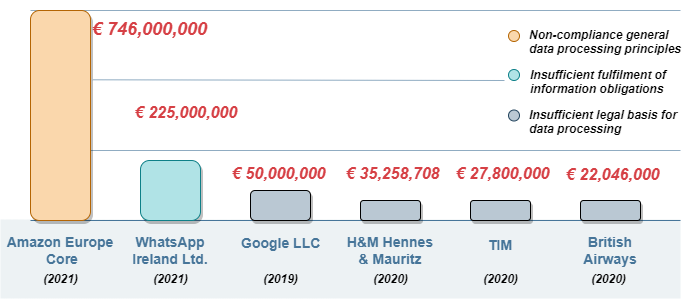} 
    \vspace{1mm}
    \caption{Major corporate fines from GDPR~\cite{finesListGdpr} for privacy issues and breaches }\vspace{-3mm}
    \label{privacy_cost}
\end{figure}

In terms of power consumption and processing, the cost of privacy can be considered inevitable when implementing B5G/6G. As techniques such as privacy-preserving algorithms and protocols carry the extra computation, it may affect any system's overall performance and energy consumption. Since we cannot fully minimize this issue, understanding the right balance between the cost and the risk is the most important aspect that can be considered here. Another issue is not adhering to proper privacy standards. Figure~\ref{privacy_cost} illustrates the top six fines imposed by GDPR on various corporates in EU~\cite{finesListGdpr} by the end of 2021. These high fines show that the costs needed to pay for the lack of sufficient improvements in privacy are very high compared to the occasional development costs needed to implement these improvements. 

The trade-off issue for implementing edge computing protocols and energy consumption is raised on~\cite{ranaweera2021survey}. The trade-off of privacy is also extended to communication when implementing protocols. The work in~\cite{singelee2011communication} shows that two critical factors for the energy cost of security and privacy are the cost of computation and communication. When protocols rely on lightweight computation, it may require a lot of communication. On the other hand, a relatively heavy computation on methods such as public-key encryption may be required, but their communication requirements are lower. The authors in~\cite{sun2019privacy} show that it is challenging to design lightweight encryption schemes for multi-authorization attributes to resist privacy leaks in the cloud. If we consider latency due to privacy,~\cite{giaconi2018privacy} shows there is a significant loss in accuracy and timely value of data in smart meters of smart grids if privacy-enhancing technique such as adding noise is done on data. 

Therefore, considering these factors, we need to create lightweight Privacy Enhancing Technologies (PET) and dynamic energy-saving methodologies to mitigate the computation drawback in B5G/6G. We propose AI-based methods to identify the right balance between privacy and cost. However, proper explanations for these methodologies should be provided to verify if the decisions are reasonable.
\vspace{-1mm}
\subsubsection{Privacy differences based on location}
Most countries should enforce legal actions and agree to cooperate to make personal data privacy regulations effective globally. However, based on location, the definitions of privacy can be different due to many reasons, such as cultural influence, religion, government regulations, beliefs, etc. Therefore, in reality, a practical challenge in solving disputes among legal entities exists. This may be challenging in B5G/6G to impose privacy preservation techniques and make use of collaborative data sharing.
We have seen a shift in power from the individual to the government in the recent past. It allowed the government to handle the privacy of individuals. Meanwhile, its activities are opaque to the general public~\cite{cavoukian2017global}. There is a discrepancy between the public safety and privacy of individuals. This is especially related to the increase of measures of protection, with the rise of terrorism and political unrest~\cite{cavoukian2017global}. Therefore, privacy measures may be customized depending on the region and on complex geopolitical scenarios that dynamically change over time. However, there are some attempts to achieve uniformity of privacy regulations in some areas of the globe. One such example is the GDPR, which was first proposed in 2012~\cite{bennett2018european}. It acts as the fundamental EU law on data protection and privacy regulations in the EU and the European Economic Area. However, it is only affecting a part of the world, and the existence of a law does not guarantee its implementation as some of these laws are only symbolic~\cite{bennett2018european}.

Another issue is specific action applicable in a country may not be valid in another country due to its nature-like demographics. The work in~\cite{cavoukian2017global} discusses a scenario of a surveillance program in Pakistan, which uses the random forest ML algorithm and reaches a false positive rate of 0.18\%, which is not applicable for more populous North America.
Therefore, these discrepancies may affect future network users when considering privacy trade-offs such as latency and cost. It will be challenging to implement such flexible privacy measures 
in B5G/6G, even with AI and XAI techniques due to their complexity and broad parameterization.
\vspace{-1mm}
\subsubsection{Difficulty in defining levels and indicators for privacy}
Privacy is a subjective concept that varies based on numerous reasons, such as personal views, culture, and geographic location. Therefore, defining to what extent users need privacy guarantees or how much privacy is violated is challenging. However, such quantification of privacy can be very useful in B5G/6G for a better-automated decision-making process in AI, as it simplifies which actions to be taken to ensure the privacy level. It will then define which steps should be taken on privacy violations quickly. Also, the entities that describe and quantify privacy should justify their quantification. Having precise levels of privacy may support the explainability of AI decisions for privacy.

Having proper privacy mechanisms ensures trust in users. However, the work in~\cite{sun2019privacy} discusses that a standardized evaluation criterion for trust needs to be improved in cloud computing, and a quantitative trust computing algorithm is required to evaluate and compare the reliability of entities accurately. Lack of privacy quantification may cause problems in ensuring guarantees for user applications.
Also,~\cite{chang2018user} describes that it needs to be clarified how to assign coefficients for precise quantification of the importance of privacy in a smart home. The authors in~\cite{bhattacharjee2021vulnerability} discuss the drawbacks of Cyber-Physical Systems (CPS) data privacy, where the authors show that having data privacy characterization and quantification model help ensure privacy preservation. Still, such models are lacking at the moment. 
These associated works show a need for such attempts to quantify privacy and violations in B5G/6G. Hence, potential methodologies for such definitions must be extensively developed, which we consider challenging due to their complexity.

\vspace{-3mm}
\subsection{Privacy Solutions for 6G Personal Data} \label{sec:solutions}

\subsubsection{Non-centralized machine learning techniques}
Recent developments in non-centralized ML techniques can proliferate the AI for B5G/6G, with their significant privacy consideration by design. One of the main problems that is encountered in traditional centralized ML is the end users need to send their data to a third-party service that host the ML model for training and predictions. Furthermore, getting better predictions from local training of AI models may not be possible with limited data, as well as performance limitations in end devices can hinder the training process. As a solution, the following decentralized ML techniques are proposed, and they can be used to train ML models that provide high-quality outputs while preserving personal data privacy:
\vspace{-2mm}
\paragraph{Federated learning} - Initially proposed in~\cite{mcmahan2017communication}, FL can be considered a key enabler for edge AI and will be used as a privacy-preserved ML technique in B5G/6G. The architecture of FL is based on multiple clients and an aggregator. In B5G/6G, clients can be IoT sensor devices, end-user devices such as mobile phones, VR headsets, smart devices, and smart vehicles. The clients initially receive an ML model, where they perform further training on these models from their own private data stored within the devices. Then, these locally trained model weights or gradients are sent to the aggregator, where the aggregator combines all the client updates. Then the new model weights or gradients are sent back to clients. This cycle repeats until the ML model within client devices achieves the desired predictability. The main advantage of this learning technique is the local data is not shared with any third party. Further, the relative cost of running ML training algorithms is lower since local device datasets are relatively low. Therefore, FL can be regarded as a privacy-preserved, lightweight, decentralized ML. Many works are proposed to enhance the features such as privacy, scalability, and efficiency in FL, which are essential for B5G/6G AI applications. For example, an efficient protocol for scalable FL for mobile devices was proposed in~\cite{bonawitz2019towards}. Work by Chen \textit{et al.}~\cite{chen2020asynchronous} presents an asynchronous online FL platform for edge devices. Therefore, FL applications can be widely adopted in B5G/6G edge AI and ML processes where personal data cannot be shared explicitly. However, numerous privacy attacks exist in FL that can potentially leak the privacy of model participants, such as inference and reconstruction attacks discussed in section~\ref{sec:issues}. The many Privacy Preserved Federated Learning (PPFL) techniques can be applied to FL to mitigate the risks of these attacks. They can be modifications to data such as perturbation, updates in AI algorithms like regularization, or enhanced model sharing like secured Multiparty Computation (MPC). 

\vspace{-2mm}
\paragraph{Secured multiparty computation-based mechanisms} - The MPC technique accompanies distinct, connected multiple parties of computing devices to jointly compute a function in a secured manner~\cite{lindell2020secure}. For this, one approach is that the parties can split the problem into small secret components and share them to solve the problem partially. The outputs of the sub-problems can be shared among all participants to obtain the final result. Here, the original data is not shared, but only partial components are shared with a third party, which is better privacy preserved than sending plain data. Therefore, when applied with ML, MPC can provide better privacy protection while maintaining model utility in a decentralized manner. MPC can be used with FL for a better privacy guarantee in FL. For example, one of the possible leakages in FL is via a malicious aggregator, which can track local model updates and get inferences on clients' private data. The work in~\cite{li2020privacy} avoids this by using a chain of clients to share local model parameters. The chained clients locally aggregate their models and send a partially aggregated model, where a malicious aggregator cannot identify individual updates. Similarly, the authors in~\cite{zhang2022augmented} propose split weights with secret shares where attackers cannot resolve a client or a server model. The advantage of MPC-based techniques in FL is the model aggregation does not cost accuracy drops, and the original model can be obtained for authorized parties if needed. However, MPC may incur communication overhead among peers and the network. 
\vspace{-2mm}
\paragraph{Split learning} - Split learning is also an emerging decentralized learning technique where clients do not fully train an ML model. Instead, they partially train a few layers in the NN and send smashed data to the server to train the remaining layers. The work in~\cite{thapa2022splitfed} combines FL with split learning, which can be regarded as a better option for resource-constrained environments since clients need lesser computation power for training.

Therefore, we identify non-centralized ML as a possible technology in B5G/6G, especially working well with resource-constrained devices in IoT environments. As discussed, several learning techniques can be combined to provide better privacy preservation. This will support delivering better edge AI and services for B5G/6G, meanwhile assuring the personal data is kept safe within the user devices.

\vspace{-2mm}
\subsubsection{Differential Privacy (DP) and other data perturbation techniques}
\label{diff_solution}
One of the main reasons to incur AI model privacy attacks is the overfitting of the models to training data\cite{zhang2020gan}. This is valid for both traditional centralized or decentralized ML techniques like FL~\cite{zhang2020gan}. If an attacker is able to recover or identify information of original data owners, it may result in significant privacy damage if the dataset consists of sensitive information. However, with perturbations added to a dataset, it will create plausible deniability to admit the data carries actual details on the owners~\cite{cormode2018privacy}. Perturbation can be done by adding noisy data or using a trained ML model to add the noise before using ML or FL model training. One key perturbation mechanism is DP, which is widely used in centralized and non-centralized learning like FL\cite{wei2020federated}. In FL, either participants or aggregators in decentralized learning can add a measurable quantity of noise to the perturb data or the model. 

\begin{figure}[ht]
    \centering
    \includegraphics[width=0.8\linewidth]{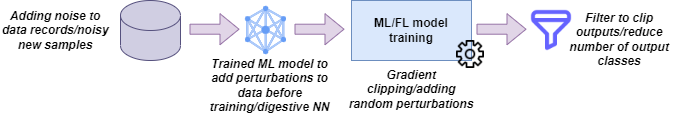} 
    \caption{Different perturbation techniques used for ML to defend against privacy attacks during the training and inference phases}\vspace{-3mm}
    \label{perturbation_mech_fl}
\end{figure}

Alternative approaches include sending data through a model that adds perturbations before using it for model training. Authors in~\cite{lee2021digestive} use a Neural network (NN) called Digestive NN to add perturbations to the original data before training. During the model training, approaches such as regularization and gradient clipping can be used to prevent the local model overfitting, which can mitigate membership inference attacks~\cite{zhang2020gan}. The ML model outputs can also be filtered in the inference phase since more output classes could leak more information for inference attacks. A similar solution is applied by the authors in~\cite{yang2020defending} for ML, which uses a prediction purification technique where a new purifier model is used to reduce the model prediction dispersion. Also, the work in~\cite{jia2019memguard} uses noise for the confidence score vector predicted by the ML target classifier. Figure~\ref{perturbation_mech_fl} provides an overview of the perturbation privacy mechanisms that can be incorporated with ML.

However, one major drawback of adding noise is the possible accuracy drop for the model predictions. Thus, the model utility may get affected by more noise input to the data. As a possible solution to this issue in FL, the work in~\cite{kim2021federated} attempts to provide a solution by adding lesser noise with better privacy to FL training. They use Local DP (LDP) with Federated Stochastic Gradient Descent (FedSGD) and compare the trade-offs between privacy and utility. They define a minor noise variance bound for FedSGD, which guarantees a required LDP level after multiple training rounds. Similarly, further research can continue reducing the impact of accuracy on ML models with perturbation, which will benefit B5G/6G privacy and lead to further industry adoption.

\vspace{-3mm}
\subsubsection{Homomorphic Encryption (HE)} 
The encryption facilitates safe transmission between two entities, such that, even if the communication is compromised, the data that is sent is unreadable to a malicious or curious third party. However, the privacy of sensitive information is at risk if they are exposed to processing with a third-party 6G-based service, even though it is safely encrypted during communication. The concept of HE can be used to address this issue. Even though previously discussed perturbations can be used when it is required to keep the data in the original form, HE can be a better option since the original data is not mutated with noise. HE is a type of encryption that allows a third party to perform computable functions on the encrypted data while preserving the encrypted format without disrupting the features of the function and format~\cite{acar2018survey}. Especially with the recently introduced fully HE, any computable function can be performed on encrypted data~\cite{acar2018survey}. The work in~\cite{khan20206g} considers HE as a key enabler for 6G wireless systems for privacy preservation in the edge/cloud. Similarly,~\cite{siriwardhana2021ai} considers HE a defense mechanism for API-based attacks against 6G systems on edge. However, they comment on further research to ensure data integrity when performing the encryption, even after performing operations on them. Limitations such as computational cost due to complexity in operations, bounded user count, and inability to perform HE on multikey encrypted data are some of the gaps to be addressed~\cite{peralta2019homomorphic,acar2018survey} in further research for B5G/6G adoption.
\vspace{-1mm}
\subsubsection{Lightweight cryptographic techniques for the IoT and edge networks}
Generic cryptographic techniques may consume a considerable computing cost, depending on the nature of the algorithm and the level of security and privacy required. Due to the lack of high computing power, not all cryptographic techniques can be practically used in IoT and edge AI environments in B5G/6G. Therefore, the identification and development of lightweight yet secure cryptographic methods on the resource-constrained IoT and the edge is another important step we identify as a solution for B5G/6G to ensure personal data privacy. Since the attackers easily target the communication channel between the IoT data generator and the edge, if the communication is less secure, the adversaries may launch man-in-the-middle and eavesdropping attacks, which can threaten privacy. The survey in~\cite{singh2017advanced} summarizes a set of techniques, including symmetric cryptographic algorithms such as AES, RC5, and TEA and asymmetric algorithms like ECC and RSA. The work in~\cite{shahidinejad2021light} proposes a lightweight authentication protocol for IoT devices in an edge-cloud environment. It uses a trust center, an intermediate between the IoT devices and the service provider cloud, responsible for coordinating the communication securely between the two. The authors in~\cite{panahi2021performance} evaluate the performance of different lightweight encryption algorithms using real IoT testbeds and show lightweight block ciphers can provide secure communication; meanwhile, multiple factors like energy and memory usage will differ based on the algorithm. Therefore, this area is open for future research to further reduce performance tradeoffs. 
\vspace{-2mm}
\subsubsection{Fog computing privacy preservation}
Fog computing is an extra layer between the edge devices and the cloud server, acting as an intermediate entity for tasks such as filtering data and forwarding them to the cloud. Such an extra layer can reduce the cost of privacy enhancements in B5G/6G since the fog layer can compute many intermediate functions. Only a fraction of data will be sent to the cloud, reducing the cloud server's overhead and network congestion. Therefore, fog computing could also help preserve the privacy of IoT and the users since they minimize the need to transmit sensitive data to the cloud for analysis~\cite{alrawais2017fog}. However, the privacy aspects within the fog node are also crucial since data from the edge will directly contact the fog layer. Also, there is a potential risk of having a compromised fog node, which the attackers may eavesdrop on or directly modify user data. Therefore, numerous works are available that propose privacy solutions for fog computing. For example,~\cite{yang2018machine} introduces a multi-functional data aggregation method with differential privacy based on ML for fog computing. The work in~\cite{qu2020decentralized} uses blockchain-enabled FL to achieve decentralized privacy, eliminate poisoning attacks, and achieve high efficiency in fog computing. The survey in~\cite{mukherjee2017security} summarizes some available techniques for privacy preservation in fog computing. It presents the following open challenges: the dynamic nature of fog nodes and users, malicious fog nodes, malicious insider attacks, secure communication, digital evidence of fog events, trust of end-users, and fog service. We see fog computing as a promising bridge between the edge and the cloud. However, there exists a possibility of being attacked by an adversary to collect data from many edge devices, causing privacy breaches. However, fog computing is a much more active area of research as far as privacy is concerned, where associated works mention privacy enhancements can be made with the fog nodes. 
\vspace{-2mm}
\subsubsection{Blockchain-based consensus and secure storage}
Blockchain is a widely used technology to verify and trace transactions done with untrustworthy parties. Blockchain can play a significant role in B5G/6G in intelligent resource management and high security and privacy features such as privacy preservation for contents, authentication, and access control, maintaining data integrity, accountability, and providing scalability on demand~\cite{hewa2020role}. Personal data can be stored in blockchain-based decentralized storage, and their ownership can be securely maintained via smart contracts~\cite{benisi2020blockchain}. Furthermore, edge AI like FL can utilize the blockchain to reach fully decentralized models without a central aggregator. Blockchain consensus can be used to aggregate the models more transparently and accountably since peers can verify the aggregated models~\cite{lu2019blockchain,li2020blockchain}. Therefore, we identify blockchain can be a prominent solution for accountability and decentralizing personal data privacy.
\vspace{-4mm}
\subsubsection{Techniques on privacy definitions and quantification}
Having proper definitions and metrics solves the issue of understanding levels of privacy and its violations. It may help solve legal disputes on privacy based on location since having precise metrics and quantification help convince many entities to have a common basis and agree on these definitions. 
Organizations may lack proper and standardized metrics for privacy. For example,~\cite{wang2018privacy} shows that service providers or individuals relax privacy measures to maintain a certain level of QoS. 
As a solution, the previously discussed DP also provides a measurable quantity of privacy by adding a desirable amount of noise to the original data. Especially $(\epsilon,\delta)$-DP consists of parameters that provide flexibility to quantify privacy~\cite{zhao2019reviewing} and tune privacy vs. utility trade-off, according to the requirements.

Often governments have less transparency on what data is collected from citizens in surveillance. The governments usually claim this is done for the safety of the public. However, if the public gets an agreement with the government administration on the levels, distrust in citizens about the government organizations collecting data may be compensated. The levels can be lifted or relaxed depending on the state's safety, but they should be defined precisely. This may be a crucial privacy step in futuristic aspects such as smart cities that are accompanied by B5G/6G, where central authorities will track massive details on user behaviour in real time.~\cite{diamantopoulou2017privacy} proposes a PLA to formalize a mutual agreement between a citizen and a Public Administration for transparency in data sharing and the privacy needs of citizens. The work in~\cite{diamantopoulou2017metamodel} proposes a metamodel for PLA to formalize the relationships of privacy-related concepts of the GDPR. However, there could be privacy disagreements between parties due to ill-defined privacy levels. This may get subjected to legal actions or discontinuation of agreements between those parties.
Providing definitions and quantification for privacy is challenging, as people's perspectives on privacy could be different based on many reasons, including personal preferences. However, in B5G/6G networks, with more opportunities to exploit user privacy, there should be proper metrics to measure privacy and provide guarantees by organizations and governmental agencies. 
\vspace{-2mm}
\section{Privacy of 6G Non-Personal Data} 
\label{sec:non-personal}




In the previous section, we mainly focused on the B5G/6G networks associated with personal data privacy issues and their solutions in previous sections. However, there is a huge collection of data belonging to the category of non-personal data, which needs attention to identify potential privacy threats, not specifically for natural persons, but also for industries and organizations that generate this data. 
\vspace{-2mm}
\subsection{Issues and Challenges for Non-Personal Data in B5G/6G}

As non-personal data is also expanding rapidly, special attention should be given to potential privacy issues that could occur with it. We see there are different issues specific to non-personal data like the ones that are described below:

\vspace{-1mm}
\subsubsection{Difficulty determining border between personal and non-personal data}
As mentioned in the introduction of non-personal data in Section~\ref{sec:data_types}, the borderline between personal and non-personal data depends on multiple variables, such as cost, time, and technological progress in the effort to recover personal data from non-personal data. It has a high likelihood that these variables will change over time. Furthermore, the definition of non-personal data is passive, implying non-personal data is anything that is not a natural person's related data. However, this definition may need a more precise nature since it assumes that personal data is easily traceable and well-defined, which is not often the case as personal data is also contingent and context-specific~\cite{somaini2020regulating}. 
Though the FFD framework specifies that FFD does not affect the legal framework to protect the privacy of natural persons \cite{nonpersonaleu}, there may not be a guarantee data shared under FFD can undergo privacy leakages future. In such a case, the framework specifies that data should be treated as personal data and required EU regulations are applied \cite{nonpersonaleu}. However, if they are shared and consist of sensitive attributes, that might already have damaged the privacy of individuals. Therefore, careful identification is necessary or otherwise, and this could cause loopholes in legal requirements. Then, it may lead to unaddressed privacy violations. 
Therefore, proper definitions and methodology for determining non-personal data should be standardized in detail, as the current approaches need improvements. 

\vspace{-3mm}
\subsubsection{Mixing personal data with non-personal data} During the data life cycle, non-personal data will be collected from various sensors simultaneously in B5G/6G associated services. At the same time, it may get coupled with related personal data, like video footage, location, or temperature changes, when the real environments involve natural persons. In such a case, processing, storage, and sharing data may be done without explicitly separating personal data from non-personal data. Several reasons, such as weaknesses in the design phase, difficulty in identifying the borderline between personal and non-personal data, and higher separation costs, may cause this issue. When coupled, it would inherently make this personal data, though it may be mistakenly considered non-personal and shared with third parties or the public.
Mixing personal and non-personal data in datasets is considered a significant challenge~\cite{graef2018towards} for non-personal data privacy. This could affect the B5G/6G networks in previously discussed areas, such as causing ambiguity in data ownership and lack of sufficient data to train models in B5G/6G networks due to difficulty in freeing non-personal data. It could create costs in separating non-personal data as well. 
However, it could be possible to avoid this problem firsthand if a clear design is made to separate personal and non-personal data during data collection, processing, or storage stages. 
\vspace{-3mm}
\subsubsection{Utility vs privacy of anonymized data}
By definition, there is no traceable, identifiable, or reachable data linked with a natural person in anonymized data\cite{kiran2022k}. Therefore, they can be considered as non-personal data. The work in~\cite{murthy2019comparative} provides a set of anonymization techniques: 1) \textit{generalization}, which replaces the value with less specific but semantically consistent values, 2) \textit{suppression} that completely remove entire parts of data such as columns or rows to make it non-personal, 3) \textit{distortion}, that introduces reversible modification to data via a function such as a hash function, 4)\textit{swapping}, which randomly rearrange variables within columns, 5) \textit{masking}, that makes an inconceivable variable via changing characters in a selected attribute. Multiple techniques can be combined to obtain more anonymity with data. The works~\cite{kiran2022k,eyupoglu2018efficient} combine perturbation techniques for anonymization of data. Similarly, synthetic data is used for anonymization in~\cite{yoon2020anonymization} by combining with actual health data records. However, with regard to anonymization approaches, the utility of data may degrade, especially with techniques such as perturbation~\cite{haoxiang2021big}, and noisy data may reduce the quality of AI model predictions. Further, information may get lost, through approaches such as suppression, that limits the predictability of models. The anonymization techniques can remove the connections and patterns among the data records, which may affect the overall utility of models trained via the data. But 6G services may require better accurate AI models with higher success rate for the proper functioning with higher QoS. Therefore, it can be considered as a significant challenge to maintain the utility tradeoff with anonymization.
\vspace{-3mm}
\subsubsection{Privacy attacks on anonymized and synthetic published data}
Even though the data is made anonymized using different techniques, there are still possible loopholes in these techniques to infer the membership of a natural person or nonhuman objects from these non-personal data by discovering correlations in the data using data mining techniques. There will be intentional or unintentional inferences from this data. Discovering a natural person's identity may affect the data subjects since the intention of anonymization was not fulfilled in such a case, thus causing a privacy leak. In 6G, the impact may affect a large population since anonymized big data generated may originally contain data related to millions of individuals and their complex connections. It may lead to revealing even further privacy-sensitive information. Only a few steps may have been taken to make data anonymized. There is no guarantee that these anonymization steps are robust to de-anonymization in the future with enhanced technology or the evolution of these attacks. The work in~\cite{majeed2020anonymization} discusses different types of privacy threats possible in Privacy Preserved Data Publishing Techniques (PPDP): 1) \textit{Identity disclosure}, where an adversary can uniquely identify an individual generally via information gained from external sources, 2) \textit{Attribute disclosure} when a sensitive attribute (e.g., criminal status, salary details, biometrics) of an individual is exposed, and 3) \textit{Membership disclosure}, where an adversary deduce the present/absent state of an individual in a given anonymized dataset with high probability.

\begin{figure}[ht]
    \centering
    \includegraphics[width=0.7\linewidth]{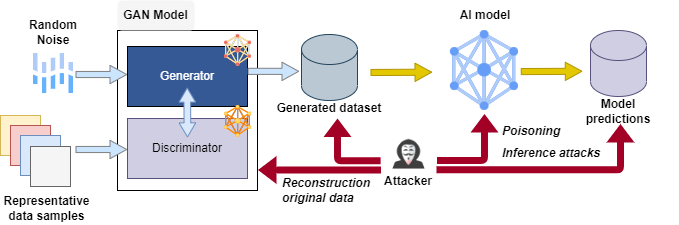} 
    \vspace{-3mm}
    \caption{GAN-based synthetic data generation and possible attacks on GAN and the data}\vspace{-1mm}
    \label{GAN_att}
\end{figure}

Many examples are available in related literature about privacy leakages from synthetic data. A possible mechanism to generate synthetic data is from generative models such as Generative Adversarial Networks (GAN)~\cite{chen2020gan}. A GAN model combines two ML models, a generator and a discriminator. Here, a GAN discriminator model should be initially trained with real data, and then it trains a generator, which will be able to generate any amount of similar synthetic data afterward. However, works~\cite{chen2020gan,el2022validating,chen2020gan} show that overfitted GAN is vulnerable to membership inference attacks since the data it generates can leak information that is available in the original overfitted dataset. An overview of possible attacks is shown in figure~\ref{GAN_att}. 
\vspace{-1mm}
\vspace{-4mm}
\subsubsection{Data localization laws vs. free flow of data}
Another issue for non-personal data comes with legal disputes among different nations. Data localization laws impose restrictions on free data movement in some countries on data of residents and data generated by corporate organizations of that country. Therefore, it clearly acts as a buffer to getting access to private data. Having free sharing capabilities for 6G non-personal data can be beneficial for organizations by reducing costs by outsourcing cheaper storage and increasing the utility of their AI models with more free data. An example of such data sharing without boundaries is the law in the EU for free flow of data~\cite{broy2017european,nonpersonaleu}. It reduces the barriers to free flow by relaxing the restrictions imposed by member states on data localization. This was done to promote a "digital single market." The framework on the free flow of non-personal data~\cite{nonpersonaleu} specifies that the data localization practices in states and vendor lock-in practices in the private sector that limits the competition, private restrictions such as legal or technical issues are obstacles to the proper functioning of data sharing. 

On the other hand, having the free flow of data can expose corporate organizations' privacy and potentially harness privacy-related information from anonymized personal data. Therefore, the privacy requirements need to be further evaluated for B5G/6G in this regard. 
Table \ref{tbl:goalsAndIssues} summarizes all the 
discussed privacy issues and their relation with the privacy goals. The issues can impact both personal and non-personal privacy goals as they can be applicable to both data types due to their borderline connectivity.
\vspace{-2mm}
\subsection{Non-Personal Data Privacy Solutions}


As earlier explained, there is difficulty in defining what non-personal data are in a strict manner.
In fact, such data may be produced either by-default, i.e., by industrial processes producing such data (e.g., machines reporting status data to back-end services), or personal data anonymized in such a way that can be considered non-personal under regulation.

In either of the two cases, the non-personal data can be protected by adversarial entities trying to infer them or poison them (or ML models trained on said data) using state-of-art methods, presented in Section~\ref{sec:solutions}, i.e., decentralized methods using secured Federated Learning or secured Multi-Party Computation, coupled with Differential Privacy or using hardware solutions such as Trusted Execution Environments, fully distributed learning methods operating in a peer-to-peer fashion, Fully Homomorphic Encryption, solutions based on blockchain or cryptographic technologies, etc.

The interesting difference between personal and non-personal data with respect to the solutions at hand is that the owner of these data becomes the key entity who must decide how to collect, store, share, process, etc., the data, and therefore, how to protect them from adversaries.

In fact, several concepts and obligations that have been defined for personal data processed in GDPR~\cite{personal_data_processing,gdprArt4,gdprArt6}, must be examined and adjusted for addressing the case of non-personal data and solutions involved with their processing:
\begin{itemize}
    \item \textbf{Privacy of entities}: Who is the entity to be protected? In some countries, organizations can be considered persons for all legal purposes.
    Therefore, an entity handling such data may need to consider how its privacy is guaranteed, along with the potential users involved in the generation of the data, if any.
    \item \textbf{Ownership}: Who owns the data? The entity owning such non-personal data can be different than the entity producing them.
    For example, if the future owner subcontracts a third-party entity to produce the data for them, or the data are produced by sensors or other devices owned by an entity other than the data receiver.
    \item \textbf{Provenance}: Who produced and/or owned specific non-personal data at a certain moment in time, and which other entities are allowed to access them?
    These questions need to be addressed, by defining and controlling specific processes and policies, to guarantee non-personal data provenance.
    \item \textbf{Storage}: Where are the data stored? Localization of non-personal data must be clearly defined to engage the appropriate regulations.
    For example, data stored in the EU can be governed by free flow~\cite{broy2017european}, but in the rest, Non-EU, world, this is unclear and important to be defined.
    \item \textbf{Sharing}: How is the sharing of non-personal data managed? Copying, sending, and overall sharing of non-personal data with other entities must be properly defined and governed by processes complying with appropriate regulations.
    \item \textbf{Processing}: Which entities are allowed to process specific non-personal data? Under what conditions? For example, data produced by entity A may be shared with another entity B, who can process them for different reasons, e.g., to produce analytics useful to entity A and/or B or even sell them to another entity C.
    Such interactions could be clearly governed by contractual agreements etc.
    \item \textbf{Anonymity}: What if entity B gets access to the non-personal data from entity A and tries to combine them with other data received from entity C to break anonymity of users who were involved in the production of data from A?
    Which entity is to blame in case of law violation?
    Also, what kind of guarantees should the non-personal data have from the point of view of anonymity (if they were produced by anonymizing personal data) that, when combined with other data, they cannot expose the original users?
    Sharing and processing purposes should strictly exclude efforts to de-anonymize users who may have been involved in the production of said data.
    \item \textbf{Legacy data}: When older non-personal data have been found to be non-compliant with respect to personal data (e.g., an entity broke their anonymization), there is the question of how should these old ``non-personal'' data be treated, if they should be deleted, or anonymized with a new process, etc.
    \item \textbf{Formatting and APIs}: There needs to be standardization of both data types and sharing processes to be able to not only facilitate and speed-up sharing and processing but also auditing of non-personal data for compliance with regulations.
\end{itemize}
All the above points are important to be defined before solutions that enable entities to process and share non-personal data can be created or customized per case.

\begin{sidewaystable}[!htbp]
\scriptsize
\centering
\caption{Privacy goals and associated issues in B5G/6G for personal and non-personal data} \label{tbl:goalsAndIssues}

\begin{tabular}{|p{1.3cm}|p{3cm}|
>{\columncolor{red!15}}c c
>{\columncolor{green!15}}c |p{5cm}|cc
>{\columncolor{green!15}}c 
>{\columncolor{green!15}}c 
>{\columncolor{green!15}}c c
>{\columncolor{green!15}}c c
>{\columncolor[HTML]{FFFFFF}}c 
>{\columncolor{green!15}}c 
>{\columncolor{yellow!20}}c |}
\hline
\multicolumn{1}{|c|}{\cellcolor[HTML]{9FC5E8}} &
  \multicolumn{1}{c|}{\cellcolor[HTML]{9FC5E8}} &
  \multicolumn{3}{c|}{\cellcolor[HTML]{9FC5E8}\textbf{Applicability}} &
  \multicolumn{1}{c|}{\cellcolor[HTML]{9FC5E8}} &
  \multicolumn{11}{c|}{\cellcolor[HTML]{9FC5E8}\textbf{Impact of privacy issues on privacy goals}} \\ \cline{3-5} \cline{7-17} 
\multicolumn{1}{|c|}{\multirow{-2}{*}{\cellcolor[HTML]{9FC5E8}\textbf{Type}}} &
  \multicolumn{1}{c|}{\multirow{-2}{*}{\cellcolor[HTML]{9FC5E8}\textbf{Privacy goal}}} &
  \multicolumn{1}{c|}{\cellcolor[HTML]{9FC5E8}\textbf{<5G}} &
  \multicolumn{1}{c|}{\cellcolor[HTML]{9FC5E8}\textbf{5G}} &
  \cellcolor[HTML]{9FC5E8}\textbf{6G} &
  \multicolumn{1}{c|}{\multirow{-2}{*}{\cellcolor[HTML]{9FC5E8}\textbf{Reasons}}} &
  \multicolumn{1}{c|}{\cellcolor[HTML]{9FC5E8}\textit{\textbf{P1}}} &
  \multicolumn{1}{c|}{\cellcolor[HTML]{9FC5E8}\textit{\textbf{P2}}} &
  \multicolumn{1}{c|}{\cellcolor[HTML]{9FC5E8}\textit{\textbf{P3}}} &
  \multicolumn{1}{c|}{\cellcolor[HTML]{9FC5E8}\textit{\textbf{P4}}} &
  \multicolumn{1}{c|}{\cellcolor[HTML]{9FC5E8}\textit{\textbf{P5}}} &
  \multicolumn{1}{c|}{\cellcolor[HTML]{9FC5E8}\textit{\textbf{P6}}} &
  \multicolumn{1}{c|}{\cellcolor[HTML]{9FC5E8}\textit{\textbf{N1}}} &
  \multicolumn{1}{c|}{\cellcolor[HTML]{9FC5E8}\textit{\textbf{N2}}} &
  \multicolumn{1}{c|}{\cellcolor[HTML]{9FC5E8}\textit{\textbf{N3}}} &
  \multicolumn{1}{c|}{\cellcolor[HTML]{9FC5E8}\textit{\textbf{N4}}} &
  \cellcolor[HTML]{9FC5E8}\textit{\textbf{N5}} \\ \hline
 &
  Ensure the personal data privacy during data generation &
  \multicolumn{1}{c|}{\cellcolor{yellow!20}L} &
  \multicolumn{1}{c|}{\cellcolor{green!15}M} &
  H &
  Data generation can be compromised by eavesdropping and privacy attacks&
  \multicolumn{1}{c|}{\cellcolor{green!15}H} &
  \multicolumn{1}{c|}{\cellcolor{green!15}H} &
  \multicolumn{1}{c|}{\cellcolor{gray!15}NA} &
  \multicolumn{1}{c|}{\cellcolor{green!15}H} &
  \multicolumn{1}{c|}{\cellcolor{green!15}H} &
  \multicolumn{1}{c|}{\cellcolor{green!15}H} &
  \multicolumn{1}{c|}{\cellcolor{yellow!20}M} &
  \multicolumn{1}{c|}{\cellcolor{yellow!20}M} &
  \multicolumn{1}{c|}{\cellcolor{red!15}L} &
  \multicolumn{1}{c|}{\cellcolor{yellow!20}M} &
  \cellcolor{green!15}H \\ \cline{2-17} 
 &
  Privacy guarantees for edge networks and edge AI &
  \multicolumn{1}{c|}{\cellcolor{red!15}L} &
  \multicolumn{1}{c|}{\cellcolor{yellow!20}M} &
  H &
  The number of connected devices increased from 9 billion in 2012 to estimated 75 billion in 2020 \cite{yu2017survey}. &
  \multicolumn{1}{c|}{\cellcolor{green!15}H} &
  \multicolumn{1}{c|}{\cellcolor{green!15}H} &
  \multicolumn{1}{c|}{\cellcolor{gray!15}NA} &
  \multicolumn{1}{c|}{\cellcolor{green!15}H} &
  \multicolumn{1}{c|}{\cellcolor{green!15}H} &
  \multicolumn{1}{c|}{\cellcolor{green!15}H} &
  \multicolumn{1}{c|}{\cellcolor{yellow!20}M} &
  \multicolumn{1}{c|}{\cellcolor{yellow!20}M} &
  \multicolumn{1}{c|}{\cellcolor{green!15}H} &
  \multicolumn{1}{c|}{\cellcolor{green!15}H} &
  \cellcolor{green!15}H \\ \cline{2-17} 
 &
  Responsibility and accountability for personal data &
  \multicolumn{1}{c|}{\cellcolor{red!15}L} &
  \multicolumn{1}{c|}{\cellcolor{yellow!20}M} &
  H &
  In 5G, leaking personal information with heterogeneity in edge, apps, vendors \cite{aleksandrova2021right} etc. &
  \multicolumn{1}{c|}{\cellcolor{yellow!20}M} &
  \multicolumn{1}{c|}{\cellcolor{yellow!20}M} &
  \multicolumn{1}{c|}{\cellcolor{green!15}H} &
  \multicolumn{1}{c|}{\cellcolor{red!15}L} &
  \multicolumn{1}{c|}{\cellcolor{green!15}H} &
  \multicolumn{1}{c|}{\cellcolor{green!15}H} &
  \multicolumn{1}{c|}{\cellcolor{green!15}H} &
  \multicolumn{1}{c|}{\cellcolor{green!15}H} &
  \multicolumn{1}{c|}{\cellcolor{yellow!20}M} &
  \multicolumn{1}{c|}{\cellcolor{green!15}H} &
  \cellcolor{green!15}H \\ \cline{2-17} 
 &
  Guarantee of erasure and rectification of personal data &
  \multicolumn{1}{c|}{\cellcolor{red!15}L} &
  \multicolumn{1}{c|}{\cellcolor{yellow!20}M} &
  H &
  With the growth of data-exploitation, the laws like GDPR aim to address control on data \cite{van2019does} in 5G. &
  \multicolumn{1}{c|}{\cellcolor{red!15}L} &
  \multicolumn{1}{c|}{\cellcolor{red!15}L} &
  \multicolumn{1}{c|}{\cellcolor{green!15}H} &
  \multicolumn{1}{c|}{\cellcolor{yellow!20}M} &
  \multicolumn{1}{c|}{\cellcolor{green!15}H} &
  \multicolumn{1}{c|}{\cellcolor{yellow!20}M} &
  \multicolumn{1}{c|}{\cellcolor{green!15}H} &
  \multicolumn{1}{c|}{\cellcolor{green!15}H} &
  \multicolumn{1}{c|}{\cellcolor{red!15}L} &
  \multicolumn{1}{c|}{\cellcolor{green!15}H} &
  M \\ \cline{2-17} 
 &
  Achieving privacy in AI-based training and processing &
  \multicolumn{1}{c|}{\cellcolor{red!15}L} &
  \multicolumn{1}{c|}{\cellcolor{yellow!20}M} &
  H &
  Services in 5G and ZSM in B5G/6G are driven by AI, making it vulnerable to privacy attacks. &
  \multicolumn{1}{c|}{\cellcolor{green!15}H} &
  \multicolumn{1}{c|}{\cellcolor{green!15}H} &
  \multicolumn{1}{c|}{\cellcolor{yellow!20}M} &
  \multicolumn{1}{c|}{\cellcolor{green!15}H} &
  \multicolumn{1}{c|}{\cellcolor{yellow!20}M} &
  \multicolumn{1}{c|}{\cellcolor{yellow!20}M} &
  \multicolumn{1}{c|}{\cellcolor{green!15}H} &
  \multicolumn{1}{c|}{\cellcolor{yellow!20}M} &
  \multicolumn{1}{c|}{\cellcolor{green!15}H} &
  \multicolumn{1}{c|}{\cellcolor{green!15}H} &
  M \\ \cline{2-17} 
 &
  Getting explanations of AI actions for privacy requirements &
  \multicolumn{1}{c|}{\cellcolor{red!15}L} &
  \multicolumn{1}{c|}{\cellcolor{yellow!20}M} &
  H &
  With fully AI driven B5G/6G, justifications needed through explanations for AI decision making. &
  \multicolumn{1}{c|}{\cellcolor{green!15}H} &
  \multicolumn{1}{c|}{\cellcolor{green!15}H} &
  \multicolumn{1}{c|}{\cellcolor{red!15}L} &
  \multicolumn{1}{c|}{\cellcolor{yellow!20}M} &
  \multicolumn{1}{c|}{\cellcolor{red!15}L} &
  \multicolumn{1}{c|}{\cellcolor{yellow!20}M} &
  \multicolumn{1}{c|}{\cellcolor{yellow!20}M} &
  \multicolumn{1}{c|}{\cellcolor{yellow!20}M} &
  \multicolumn{1}{c|}{\cellcolor{green!15}H} &
  \multicolumn{1}{c|}{\cellcolor{yellow!20}M} &
  M \\ \cline{2-17} 
\multirow{-7}{*}{Personal data} &
  Balance interests in privacy protection in global context &
  \multicolumn{1}{c|}{\cellcolor{red!15}L} &
  \multicolumn{1}{c|}{\cellcolor{red!15}L} &
  H &
  Not every state agree on unified approach. No global mechanisms for privacy yet, existing ones like human rights are insufficient alone \cite{rojszczak2020does}. &
  \multicolumn{1}{c|}{\cellcolor{red!15}L} &
  \multicolumn{1}{c|}{\cellcolor{red!15}L} &
  \multicolumn{1}{c|}{\cellcolor{green!15}H} &
  \multicolumn{1}{c|}{\cellcolor{yellow!20}M} &
  \multicolumn{1}{c|}{\cellcolor{green!15}H} &
  \multicolumn{1}{c|}{\cellcolor{green!15}H} &
  \multicolumn{1}{c|}{\cellcolor{green!15}H} &
  \multicolumn{1}{c|}{\cellcolor{green!15}H} &
  \multicolumn{1}{c|}{\cellcolor{red!15}L} &
  \multicolumn{1}{c|}{\cellcolor{yellow!20}M} &
  \cellcolor{green!15}H \\ \hline
 &
  Identification and measurable distinction non-personal data &
  \multicolumn{1}{c|}{\cellcolor{red!15}L} &
  \multicolumn{1}{c|}{\cellcolor{red!15}L} &
  H &
  The notion of non-personal data is relatively new and unclear for most contexts. &
  \multicolumn{1}{c|}{\cellcolor{yellow!20}M} &
  \multicolumn{1}{c|}{\cellcolor{yellow!20}M} &
  \multicolumn{1}{c|}{\cellcolor{green!15}H} &
  \multicolumn{1}{c|}{\cellcolor{green!15}H} &
  \multicolumn{1}{c|}{\cellcolor{yellow!20}M} &
  \multicolumn{1}{c|}{\cellcolor{green!15}H} &
  \multicolumn{1}{c|}{\cellcolor{green!15}H} &
  \multicolumn{1}{c|}{\cellcolor{green!15}H} &
  \multicolumn{1}{c|}{\cellcolor{green!15}H} &
  \multicolumn{1}{c|}{\cellcolor{green!15}H} &
  M \\ \cline{2-17} 
 &
  Evaluation of privacy leakages from non-personal data &
  \multicolumn{1}{c|}{\cellcolor{red!15}L} &
  \multicolumn{1}{c|}{\cellcolor{red!15}L} &
  H &
  Non-personal data can be attacked for extracting privacy sensitive data in B5G/6G. &
  \multicolumn{1}{c|}{\cellcolor{yellow!20}M} &
  \multicolumn{1}{c|}{\cellcolor{yellow!20}M} &
  \multicolumn{1}{c|}{\cellcolor{green!15}H} &
  \multicolumn{1}{c|}{\cellcolor{green!15}H} &
  \multicolumn{1}{c|}{\cellcolor{green!15}H} &
  \multicolumn{1}{c|}{\cellcolor{green!15}H} &
  \multicolumn{1}{c|}{\cellcolor{green!15}H} &
  \multicolumn{1}{c|}{\cellcolor{green!15}H} &
  \multicolumn{1}{c|}{\cellcolor{green!15}H} &
  \multicolumn{1}{c|}{\cellcolor{green!15}H} &
  M \\ \cline{2-17} 
 &
  Privacy protected AI automated network management &
  \multicolumn{1}{c|}{\cellcolor{red!15}L} &
  \multicolumn{1}{c|}{\cellcolor{red!15}L} &
  H &
  AI management operations generate huge quantity of non-personal data on states and actions performed in B5G/6G. &
  \multicolumn{1}{c|}{\cellcolor{green!15}H} &
  \multicolumn{1}{c|}{\cellcolor{green!15}H} &
  \multicolumn{1}{c|}{\cellcolor{red!15}L} &
  \multicolumn{1}{c|}{\cellcolor{green!15}H} &
  \multicolumn{1}{c|}{\cellcolor{yellow!20}M} &
  \multicolumn{1}{c|}{\cellcolor{yellow!20}M} &
  \multicolumn{1}{c|}{\cellcolor{yellow!20}M} &
  \multicolumn{1}{c|}{\cellcolor{yellow!20}M} &
  \multicolumn{1}{c|}{\cellcolor{yellow!20}M} &
  \multicolumn{1}{c|}{\cellcolor{green!15}H} &
  M \\ \cline{2-17} 
\multirow{-4}{*}{\begin{tabular}[c]{@{}l@{}}Non-personal \\data\end{tabular}} &
  Standardization of privacy in technologies, and applications &
  \multicolumn{1}{c|}{\cellcolor{red!15}L} &
  \multicolumn{1}{c|}{\cellcolor{yellow!20}M} &
  H &
  No universal privacy standards, but industrialized states attempt unified data control practices \cite{rustad2019towards}. &
  \multicolumn{1}{c|}{\cellcolor{red!15}L} &
  \multicolumn{1}{c|}{\cellcolor{red!15}L} &
  \multicolumn{1}{c|}{\cellcolor{green!15}H} &
  \multicolumn{1}{c|}{\cellcolor{yellow!20}M} &
  \multicolumn{1}{c|}{\cellcolor{green!15}H} &
  \multicolumn{1}{c|}{\cellcolor{green!15}H} &
  \multicolumn{1}{c|}{\cellcolor{green!15}H} &
  \multicolumn{1}{c|}{\cellcolor{green!15}H} &
  \multicolumn{1}{c|}{\cellcolor{green!15}H} &
  \multicolumn{1}{c|}{\cellcolor{yellow!20}M} &
  \cellcolor{green!15}H \\ \hline
 &
  Ensure privacy protection for big data systems &
  \multicolumn{1}{c|}{\cellcolor{yellow!20}M} &
  \multicolumn{1}{c|}{\cellcolor{green!15}H} &
  H &
  Big data concepts were appearing in early 2000s with 3V concepts \cite{laney20013d} from 3G era. &
  \multicolumn{1}{c|}{\cellcolor{green!15}H} &
  \multicolumn{1}{c|}{\cellcolor{red!15}L} &
  \multicolumn{1}{c|}{\cellcolor{green!15}H} &
  \multicolumn{1}{c|}{\cellcolor{green!15}H} &
  \multicolumn{1}{c|}{\cellcolor{green!15}H} &
  \multicolumn{1}{c|}{\cellcolor{yellow!20}M} &
  \multicolumn{1}{c|}{\cellcolor{green!15}H} &
  \multicolumn{1}{c|}{\cellcolor{green!15}H} &
  \multicolumn{1}{c|}{\cellcolor{green!15}H} &
  \multicolumn{1}{c|}{\cellcolor{yellow!20}M} &
  M \\ \cline{2-17} 
 &
  Balance between privacy and performance of services &
  \multicolumn{1}{c|}{\cellcolor{red!15}L} &
  \multicolumn{1}{c|}{\cellcolor{yellow!20}M} &
  H &
  A growth of related works on IoT device \cite{yu2017survey}, and big data privacy \cite{yu2016big} seen since 4G/5G. &
  \multicolumn{1}{c|}{\cellcolor{red!15}L} &
  \multicolumn{1}{c|}{\cellcolor{green!15}H} &
  \multicolumn{1}{c|}{\cellcolor{green!15}H} &
  \multicolumn{1}{c|}{\cellcolor{green!15}H} &
  \multicolumn{1}{c|}{\cellcolor{yellow!20}M} &
  \multicolumn{1}{c|}{\cellcolor{yellow!20}M} &
  \multicolumn{1}{c|}{\cellcolor{green!15}H} &
  \multicolumn{1}{c|}{\cellcolor{yellow!20}M} &
  \multicolumn{1}{c|}{\cellcolor{green!15}H} &
  \multicolumn{1}{c|}{\cellcolor{green!15}H} &
  M \\ \cline{2-17} 
 &
  Achieving proper utilization of interoperability and data portability &
  \multicolumn{1}{c|}{\cellcolor{red!15}L} &
  \multicolumn{1}{c|}{\cellcolor{yellow!20}M} &
  H &
  Recent collaborations like DTP \cite{egan2019data} relevant to current 5G, further enhance in B5G/6G for cases like unified digital identity \cite{rivera2017digital}. &
  \multicolumn{1}{c|}{\cellcolor{red!15}L} &
  \multicolumn{1}{c|}{\cellcolor{red!15}L} &
  \multicolumn{1}{c|}{\cellcolor{green!15}H} &
  \multicolumn{1}{c|}{\cellcolor{green!15}H} &
  \multicolumn{1}{c|}{\cellcolor{green!15}H} &
  \multicolumn{1}{c|}{\cellcolor{yellow!20}M} &
  \multicolumn{1}{c|}{\cellcolor{green!15}H} &
  \multicolumn{1}{c|}{\cellcolor{green!15}H} &
  \multicolumn{1}{c|}{\cellcolor{green!15}H} &
  \multicolumn{1}{c|}{\cellcolor{green!15}H} &
  \cellcolor{green!15}H \\ \cline{2-17} 
\multirow{-4}{*}{Hybrid} &
  Quantifying privacy and privacy violations &
  \multicolumn{1}{c|}{\cellcolor{red!15}L} &
  \multicolumn{1}{c|}{\cellcolor{red!15}L} &
  H &
  Works \cite{chang2018user,bhattacharjee2021vulnerability} discuss on lack of privacy quantification. Privacy quantification will support B5G/6G for fully automated decision making. &
  \multicolumn{1}{c|}{\cellcolor{green!15}H} &
  \multicolumn{1}{c|}{\cellcolor{green!15}H} &
  \multicolumn{1}{c|}{\cellcolor{green!15}H} &
  \multicolumn{1}{c|}{\cellcolor{green!15}H} &
  \multicolumn{1}{c|}{\cellcolor{yellow!20}M} &
  \multicolumn{1}{c|}{\cellcolor{green!15}H} &
  \multicolumn{1}{c|}{\cellcolor{green!15}H} &
  \multicolumn{1}{c|}{\cellcolor{yellow!20}M} &
  \multicolumn{1}{c|}{\cellcolor{yellow!20}M} &
  \multicolumn{1}{c|}{\cellcolor{yellow!20}M} &
  M \\ \hline
\end{tabular}
\begin{tabular}{ll}
\begin{tabular}[c]{@{}l@{}l@{}}
\\P1 - Privacy attacks on AI models and private data
\\P2 - IoT edge network and edge AI privacy attacks
\\P3 - Privacy limitations in cloud computing and storage environments
\\P4 - Cost on privacy enhancements
\\P5 - Privacy differences based on location
\\P6 - Difficulty in defining levels and indicators for privacy
\end{tabular}
\begin{tabular}[c]{@{}l@{}}
\\N1 - Difficulty determining border between personal and non-personal data.
\\N2 - Mixing Personal Data with Non-Personal Data
\\N3 - Utility vs privacy of anonymized data
\\N4 - Privacy attacks on anonymized and synthetic published data
\\N5 - Data Localization Laws vs. free flow of data
\end{tabular}

\end{tabular}

\begin{flushleft}
\begin{center}
    
\begin{tikzpicture}
\node (rect) at (0,2) [draw,thick,minimum width=1cm,minimum height=0.7cm, fill= gray!15, label=0:Not Applicable] {NA};
\node (rect) at (4,2) [draw,thick,minimum width=1cm,minimum height=0.7cm, fill= red!15, label=0:Low Applicability] {L};
\node (rect) at (8.3,2) [draw,thick,minimum width=1cm,minimum height=0.7cm, fill= yellow!20, label=0:Medium Applicability] {M};
\node (rect) at (12.8,2) [draw,thick,minimum width=1cm,minimum height=0.7cm, fill= green!15, label=0:High Applicability] {H};
\end{tikzpicture}
\end{center}

\end{flushleft}
\end{sidewaystable}

\vspace{-1mm}
\section{Lessons Learned and Future Research Directions} \label{sec:lesson}

This section discusses the outcomes and lessons from the discussions regarding the privacy of personal and non-personal data that will be utilized in B5G/6G networks. Privacy in B5G/6G is at its early vision phase, and we consider it vital to continue this discussion relevant to data privacy, properly distinguish data types, identify potential privacy threats based on these types, and possible solutions for mitigating them. Further research questions regarding our findings and possible future approaches that can be taken are also discussed within the scope of data privacy.


\vspace{-1mm}
\subsection{6G Privacy Taxonomy and Goals}
\subsubsection{Lessons learned}
There are multiple definitions, classifications, and data types we observed regarding the privacy of data in the related literature. Privacy itself can be regarded as subjective, which can differ based on the perspective of the data owner and their preferences. Different actions on the data cycle, from generation to sharing to third parties, lead to different privacy requirements being fulfilled. In B5G/6G, clear consideration should be given to different taxonomies of privacy based on multiple perspectives of consideration on data, valid for custom preferences. 

Based on different areas considered under B5G/6G, the goals laid the foundation for the underlying privacy issues and potential solutions for personal and non-personal data. However, the difficulty of achieving these goals may not necessarily be depicted by the number of associated issues. We see some areas, such as big data and AI privacy, have made significant progress over the recent years, as suggested by the related work. We also summarized the impact of these privacy goals in Table~\ref{tbl:goalsAndIssues}, where we see pre-5G era has a lesser impact on some issues simply due to the lack of technological background required to fulfill these goals. However, the 5G and B5G/6G networks will have greater applicability to these goals. Personal data privacy issues have an impact on both personal and non-personal data since non-personal data may have unseen relationships with individuals. Similarly, the opposite can also be true, where non-personal data-related issues can affect personal data privacy goals. Therefore, these personal data issues may also affect the specified privacy goals related to non-personal data. For example, the goal of having a clear distinction between personal and non-personal data can be influenced by issues related to personal data, such as privacy attacks and costs of privacy protection. We further show this impact on privacy preservation for privacy goals in Table \ref{tbl:goalsAndIssues}.
\vspace{-1mm}
\subsubsection{Remaining research questions}
The taxonomy and privacy goals defined in our survey suggested that the following issues may need to be addressed:
\begin{itemize}
     \item How is privacy classified, and what mechanisms to data should be done based on different actions performed by non-human entities such as AI in B5G/6G?
     \item How to evaluate the difficulty level when achieving the privacy goals for different scenarios in B5G/6G service implementation, since some goals may be relatively easier to achieve than others depending on the use cases and factors such as privacy laws, location, user preferences, etc.?
\end{itemize}
\vspace{-1mm}
\subsubsection{Possible future directions}
The current landscape of B5G/6G is in its initial phase of vision and discussion state. Privacy for personal and non-personal data can be further discussed in this phase since most tasks in B5G/6G will be beyond human supervision and controlled by zero-touch networks powered by AI decision-making. Taxonomy of privacy regarding data handling should be taken into account whenever data-driven operations are designed and implemented under 6G networks. When considering the achievement of privacy goals,  we may need to carefully consider the associated issues, their severity, and appropriate privacy solutions based on the nature of the data. 

\vspace{-2mm}
\subsection{Personal Data}
\subsubsection{Lessons learned}
Any data is regarded as personal data if it has a linkage to a natural person. We observe that B5G/6G can potentially be flooded with new types of personal data via a multitude of IoT sensors and tracking devices that will be abundant with B5G/6G networks. Furthermore, with the expansion of technology, novel approaches to data collection methods will be introduced that have the capability to capture private and sensitive data from individuals. With more data collected, more accurate AI models can be made and rapidly applied in a zero-touch manner in B5G/6G service layers. However, many issues in personal data, such as privacy attacks, vulnerabilities, computing resource limitations, costs, and legal disputes, may need to be carefully considered and sufficiently addressed while designing the expected fully autonomous B5G/6G networks. 
\vspace{-1mm}
\subsubsection{Remaining research questions}
The following questions are open to research when considering personal data privacy in B5G/6G networks:
\begin{itemize}
    \item How the trade-offs among privacy, performance, costs, and utility can be balanced when accounting for privacy preservation methods for personal data?
    \item What measures can be implemented upon differences among individuals, organizations, and states when working cooperatively and sharing personal data with varying privacy requirements?
    \item How to identify unintended exploitation of personal data in advance by novel tools used to enhance interpretation and explanation of data and AI applications.
\end{itemize}
\vspace{-2mm}
\subsubsection{Possible future directions}
Therefore, possible future research can include low latency, lightweight identification mechanisms of personal data via suitable AI models. The assessment of threats before sharing data with a third party can be done. When considering new data types that are introduced with B5G/6G, careful investigation of user privacy should be made essential before adopting it for general use. Existing data may also need new mechanisms to resolve unseen connections among individuals that can potentially threaten privacy.
\vspace{-1mm}
\subsection{Non-Personal Data}
\subsubsection{Lessons learned}
We can observe that most privacy considerations and techniques are related to personal data, where the privacy of non-personal data is often under the radar. There may not be an absolute "clear-cut" between personal and non-personal data. The relatively less associated work on non-personal data privacy is more evident. However, more and more surrounding issues such as attribute privacy concerns~\cite{zhang2020attribute}, which can be incorporated with non-personal data, could cause the revealing of sensitive details of an organization, individuals, or other private data through specific attributes by non-personal data generators. Therefore, it is essential to consider this privacy aspect in non-personal data for B5G/6G networks. This is because data generated by modes such as non-human AI-based automated system logs, industrial sensor data, and many other interconnected systems can produce and communicate non-personal data at a very large scale due to the fast data transmission capabilities in B5G/6G.
\vspace{-1mm}
\subsubsection{Remaining research questions}
 The non-personal data considerations have the following questions to be further clarified:
\begin{itemize}
    \item A measurable strategy for differentiating the privacy of non-personal data from personal data is essential in quantitatively determining the risk and levels of privacy required.
    \item Definition of further refined classification of non-personal data since anonymization techniques also has different approaches to making personal data non-personal.
    \item Definition of a standard set of anonymization techniques and criteria to create non-personal data.
    \item Identification of potential privacy threats and attack scenarios for non-human data generators, such as automated system logs, often used in the industry.
\end{itemize}
\vspace{-1mm}
\subsubsection{Possible future directions}

In future work, we suggest having further investigations on non-personal data definitions and privacy considerations for multi-dimensional and different types of non-personal data. For anonymization, the work in~\cite{pawar2018anonymization} suggests new techniques of anonymization schemes universally applicable to any type of data. They could be used to generate non-personal data. Also, when considering sensitive non-human data generation, further work on isolation techniques could be used to preserve this data from attacks.

\vspace{-1mm}
\subsection{Privacy Solutions}
\subsubsection{Lessons learned}
We identified multiple privacy solutions for personal and non-personal data in B5G/6G. They are at different stages of development in the case of privacy, as some solutions use a privacy-by-design approach while some may not. Enablers such as FL and decentralized edge AI techniques can further enhance the privacy of individuals; however, these techniques themselves have privacy problems that are yet to be solved. With privacy preservation, there is a utility vs. privacy trade-off that can be expected due to loss of information due to privacy preservation. Table \ref{tbl:issuesVsSolutions} summarizes the applicability of the proposed solutions in both personal and non-personal data related issues.

\begin{table}[ht]
\scriptsize
\centering
\vspace{-2mm}
\caption{Privacy solutions and their applicability in addressing the issues in personal and non-personal data} \label{tbl:issuesVsSolutions}
\vspace{-3mm}
\begin{tabular}{|l|ccccccccccc|}
\hline
\rowcolor[HTML]{9FC5E8} 
\multicolumn{1}{|c|}{\cellcolor[HTML]{9FC5E8}} &
  \multicolumn{11}{c|}{\cellcolor[HTML]{9FC5E8}\textbf{Addressed privacy issues}} \\ \cline{2-12} 
\rowcolor[HTML]{9FC5E8} 
\multicolumn{1}{|c|}{\multirow{-2}{*}{\cellcolor[HTML]{9FC5E8}\textbf{Privacy solution}}} &
  \multicolumn{1}{l|}{\cellcolor[HTML]{9FC5E8}\textit{\textbf{P1}}} &
  \multicolumn{1}{l|}{\cellcolor[HTML]{9FC5E8}\textit{\textbf{P2}}} &
  \multicolumn{1}{l|}{\cellcolor[HTML]{9FC5E8}\textit{\textbf{P3}}} &
  \multicolumn{1}{l|}{\cellcolor[HTML]{9FC5E8}\textit{\textbf{P4}}} &
  \multicolumn{1}{l|}{\cellcolor[HTML]{9FC5E8}\textit{\textbf{P5}}} &
  \multicolumn{1}{l|}{\cellcolor[HTML]{9FC5E8}\textit{\textbf{P6}}} &
  \multicolumn{1}{l|}{\cellcolor[HTML]{9FC5E8}\textit{\textbf{N1}}} &
  \multicolumn{1}{l|}{\cellcolor[HTML]{9FC5E8}\textit{\textbf{N2}}} &
  \multicolumn{1}{l|}{\cellcolor[HTML]{9FC5E8}\textit{\textbf{N3}}} &
  \multicolumn{1}{l|}{\cellcolor[HTML]{9FC5E8}\textit{\textbf{N4}}} &
  \multicolumn{1}{l|}{\cellcolor[HTML]{9FC5E8}\textit{\textbf{N5}}} \\ \hline
Non-centralized ML &
  \multicolumn{1}{c|}{\checkmark} &
  \multicolumn{1}{c|}{\checkmark} &
  \multicolumn{1}{c|}{\checkmark} &
  \multicolumn{1}{c|}{\checkmark} &
  \multicolumn{1}{c|}{\checkmark} &
  \multicolumn{1}{c|}{} &
  \multicolumn{1}{c|}{} &
  \multicolumn{1}{c|}{\checkmark} &
  \multicolumn{1}{c|}{\checkmark} &
  \multicolumn{1}{c|}{\checkmark} &
  \checkmark \\ \hline
DP and data perturbation &
  \multicolumn{1}{c|}{\checkmark} &
  \multicolumn{1}{c|}{\checkmark} &
  \multicolumn{1}{c|}{\checkmark} &
  \multicolumn{1}{c|}{} &
  \multicolumn{1}{c|}{\checkmark} &
  \multicolumn{1}{c|}{\checkmark} &
  \multicolumn{1}{c|}{\checkmark} &
  \multicolumn{1}{c|}{\checkmark} &
  \multicolumn{1}{c|}{} &
  \multicolumn{1}{c|}{\checkmark} &
  \checkmark \\ \hline
Homomorphic Encryption &
  \multicolumn{1}{c|}{\checkmark} &
  \multicolumn{1}{c|}{\checkmark} &
  \multicolumn{1}{c|}{\checkmark} &
  \multicolumn{1}{c|}{} &
  \multicolumn{1}{c|}{\checkmark} &
  \multicolumn{1}{c|}{} &
  \multicolumn{1}{c|}{} &
  \multicolumn{1}{c|}{\checkmark} &
  \multicolumn{1}{c|}{\checkmark} &
  \multicolumn{1}{c|}{\checkmark} &
  \checkmark \\ \hline
Lightweight cryptiography for IoT and edge &
  \multicolumn{1}{c|}{\checkmark} &
  \multicolumn{1}{c|}{\checkmark} &
  \multicolumn{1}{c|}{} &
  \multicolumn{1}{c|}{\checkmark} &
  \multicolumn{1}{c|}{\checkmark} &
  \multicolumn{1}{c|}{} &
  \multicolumn{1}{c|}{} &
  \multicolumn{1}{c|}{} &
  \multicolumn{1}{c|}{\checkmark} &
  \multicolumn{1}{c|}{\checkmark} &
  \checkmark \\ \hline
Fog computing privacy preservation &
  \multicolumn{1}{c|}{\checkmark} &
  \multicolumn{1}{c|}{} &
  \multicolumn{1}{c|}{\checkmark} &
  \multicolumn{1}{c|}{\checkmark} &
  \multicolumn{1}{c|}{\checkmark} &
  \multicolumn{1}{c|}{} &
  \multicolumn{1}{c|}{} &
  \multicolumn{1}{c|}{} &
  \multicolumn{1}{c|}{\checkmark} &
  \multicolumn{1}{c|}{\checkmark} &
  \checkmark \\ \hline
Blockchain-based consensus and  storage &
  \multicolumn{1}{c|}{\checkmark} &
  \multicolumn{1}{c|}{\checkmark} &
  \multicolumn{1}{c|}{\checkmark} &
  \multicolumn{1}{c|}{\checkmark} &
  \multicolumn{1}{c|}{\checkmark} &
  \multicolumn{1}{c|}{} &
  \multicolumn{1}{c|}{} &
  \multicolumn{1}{c|}{\checkmark} &
  \multicolumn{1}{c|}{\checkmark} &
  \multicolumn{1}{c|}{\checkmark} &
  \checkmark \\ \hline
Privacy levels definitions and quantification &
  \multicolumn{1}{c|}{} &
  \multicolumn{1}{c|}{} &
  \multicolumn{1}{c|}{\checkmark} &
  \multicolumn{1}{c|}{\checkmark} &
  \multicolumn{1}{c|}{\checkmark} &
  \multicolumn{1}{c|}{\checkmark} &
  \multicolumn{1}{c|}{\checkmark} &
  \multicolumn{1}{c|}{\checkmark} &
  \multicolumn{1}{c|}{\checkmark} &
  \multicolumn{1}{c|}{} &
  \checkmark \\ \hline
\end{tabular}
\begin{tabular}{ll}
\begin{tabular}[c]{@{}l@{}}\\P1 - Privacy attacks on AI models and private data
\\P2 - IoT edge network and edge AI privacy attacks
\\P3 - Privacy limitations in cloud computing and storage environments
\\P4 - Cost on privacy enhancements
\\P5 - Privacy differences based on location
\\P6 - Difficulty in defining levels and indicators for privacy
\\ \end{tabular} & \begin{tabular}[c]{@{}l@{}} 
\\N1 - Difficulty determining border between personal \\and non-personal data.
\\N2 - Mixing Personal Data with Non-Personal Data
\\N3 - Utility vs privacy of anonymized data
\\N4 - Privacy attacks anonymized and synthetic data
\\N5 - Data Localization Laws vs. free flow of data\end{tabular}
\vspace{-3mm}
\end{tabular}
\end{table}

\vspace{-2mm}
\subsubsection{Remaining research questions} 
The following questions are identified as potential issues that are still available in privacy solutions in general:
\begin{itemize}
\item How to efficiently combine multiple privacy preservation techniques in a framework for B5G/6G oriented services?
\item What options are available for privacy solutions to be applied in existing platforms that do not natively build with privacy consideration?
\item What further metrics can be discovered to quantitatively evaluate privacy enhancement and utility with single or multiple privacy solutions?
\item How can these solutions impact the overall quality of services in the short and long terms?
\end{itemize}
\vspace{-1mm}
\subsubsection{Possible future directions}
Therefore, we propose that future directions should include a clear mechanism for defining which privacy solution(s) should be imposed upon a careful assessment of potential privacy leakages and vulnerabilities to both personal and non-personal data in data managing frameworks of B5G/6G. The general framework for combined privacy preservation techniques can also be proposed as a standardization approach for privacy in B5G/6G.
\vspace{-1mm}
\section{Conclusion}\label{conclusion}
In this survey, we have comprehensively discussed the privacy of personal and non-personal data for B5G/6G networks. Non-personal data is differentiated from personal data based on the likelihood of resolving identifiable information of a natural person from data. Thus, the data is determined based on the data subject. However, when considering the probability for 6G, a decade of future potential should be accounted for the technological progress, as seemingly non-personal data may have an indirect relationship with an individual that may be possible to trace with enhanced capabilities in computing and AI. We presented a set of privacy goals to be achieved in B5G/6G by considering the current gaps identified in each category. Both personal and non-personal data in B5G/6G are potentially vulnerable to numerous privacy attacks and threats. The contributing factors include limitations in hardware and network resources, vulnerabilities in AI models, higher cost of implementation, changing laws, and ambiguity in differentiating privacy types. As solutions, we proposed existing and novel techniques used to solve privacy problems. However, the proposed solutions themselves may need to be completed, thus leaving future directions that can be taken to safeguard privacy for future networks.
\vspace{-1mm}
\section{Acknowledgments}

This work is partly supported by European Union in SPATIAL  (Grant No: 101021808), Academy of Finland in 6Genesis (grant no. 318927) and  Science Foundation Ireland under CONNECT phase 2 (Grant no. 13/RC/2077\_P2) projects.

\bibliographystyle{unsrt}
\bibliography{bibliography}

\appendix









\end{document}